\documentclass[aps,pra,twocolumn,longbibliography,superscriptaddress]{revtex4-2}

\usepackage[T1]{fontenc}
\usepackage[utf8]{inputenc}
\usepackage{color}
\usepackage{babel}
\usepackage{amsmath}
\usepackage{amssymb}
\usepackage{subfigure}
\usepackage{graphicx}
\usepackage{physics} 
\usepackage{dsfont}
\usepackage{hyperref}
\usepackage{cleveref}
\usepackage{svg}
\usepackage{subcaption}
\usepackage{epsfig}
\usepackage{ragged2e}
\usepackage[normalem]{ulem}
\usepackage{comment}
\usepackage{booktabs}
\usepackage{comment}

\newcommand{\zanin}[1]{{\color{orange} #1}}

\begin{document}

\title{Structured light under turbulence}

\author{Guilherme S. Barros \href{https://orcid.org/0009-0003-0471-0536}{\includegraphics[scale=0.05]{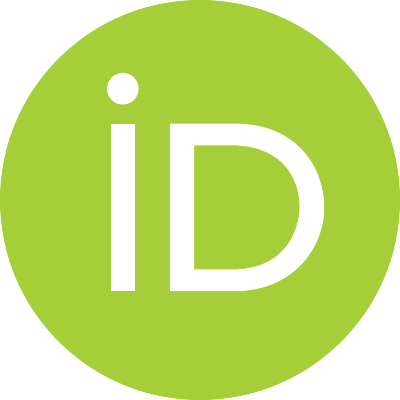}}}
\email{}
\affiliation{Institute of Physics, Federal University of Goi\'as, Goi\^ania, Goi\'as, 74.690-900, Brazil}

\author{Lucas C. C\'eleri\href{https://orcid.org/0000-0001-5120-8176}{\includegraphics[scale=0.05]{orcidid.pdf}}}
\email{lucas@qpequi.com}
\affiliation{Institute of Physics, Federal University of Goi\'as, Goi\^ania, Goi\'as, 74.690-900, Brazil}
\affiliation{Instituto de Física de São Carlos, Universidade de São Paulo, CP 369, 13560-970, São Carlos, SP, Brasil}

\author{Antonio Zelaquett Khoury\href{https://orcid.org/0000-0002-7487-5067}{\includegraphics[scale=0.05]{orcidid.pdf}}}
\email{azkhoury@id.uff.br}
\affiliation{Instituto de Física, Universidade Federal Fluminense, Niteroi, RJ, 24210-346, Brazil}

\author{André L. S. Santos Junior\href{https://orcid.org/0000-0002-2664-0781}{\includegraphics[scale=0.05]{orcidid.pdf}}}
%\email{azkhoury@id.uff.br}
\affiliation{Instituto de Física, Universidade Federal Fluminense, Niteroi, RJ, 24210-346, Brazil}

\author{Rafael M. Gomes\href{https://orcid.org/}{\includegraphics[scale=0.05]{orcidid.pdf}}}
\affiliation{Institute of Physics, Federal University of Goi\'as, Goi\^ania, Goi\'as, 74.690-900, Brazil}

\author{Guilherme L. Zanin\href{https://orcid.org/0000-0003-4178-7818}{\includegraphics[scale=0.05]{orcidid.pdf}}}
\email{guilherme_zanin@ufg.br}
\affiliation{Institute of Physics, Federal University of Goi\'as, Goi\^ania, Goi\'as, 74.690-900, Brazil}

\begin{abstract}
Structured light has emerged as a promising resource for high-capacity and secure free-space optical communication, where atmospheric turbulence remains a major source of signal degradation. In this work, we investigate the resilience of different transverse mode structures of an optical beam with respect to the random action of turbulence. A spatial light modulator (SLM) is programmed to apply amplitude and phase modulation corresponding to the desired transverse mode, which is then transmitted through a turbulent medium emulated by a second SLM. Laguerre–Gauss, Hermite–Gauss, and Airy beams are investigated through the resulting intensity distributions measured with a camera. Different figures of merit are used to evaluate and compare the resilience of these modes.
\end{abstract}

\maketitle

%%%%%%%%%%%%%%%%%%%%%%%%%%%%%%%%%%%%%%%%%%%%%%%%%%%%%%%%%%%%%%%%
%%%%%%%%%%%%%%%%%%%%%%%%%%%%%%%%%%%%%%%%%%%%%%%%%%%%%%%%%%%%%%%%
%%%%%%%%%%%%%%%%%%%%%%%%%%%%%%%%%%%%%%%%%%%%%%%%%%%%%%%%%%%%%%%%

\section{Introduction}

The need for secure communication has driven technological evolution from ancient civilizations to the present day. Early methods relied on simple codifications, such as the Caesar cipher, which later evolved into sophisticated cryptographic machines like the Enigma during World War II, and eventually into the RSA algorithm~\cite{rsa1978}, which remains widely utilized in modern classical networks~\cite{Stinson2023,Menezes1996}. More recently, this trajectory has culminated in the development of quantum communication protocols. Notably, the seminal BB84 protocol pioneered the use of photon polarization as a degree of freedom~\cite{BB84}, while the E91 protocol introduced the concept of entanglement-based quantum cryptography~\cite{E91}. See Refs.~\cite{Gisin2002,Scarani2009} for reviews on quantum cryptography and quantum key distribution.

Currently, global data transmission relies heavily on optical fiber communication systems, where digital information is transmitted through sequences of temporally modulated optical pulses carrying encrypted binary data~\cite{Agrawal2002}. However, the electromagnetic field possesses a rich variety of degrees of freedom that can be exploited to encode information, including polarization, time-bin, frequency, quadratures, and transverse spatial modes, which have become key resources for both classical and quantum photonic information processing~\cite{Flamini2019}. Particularly relevant to this work are the transverse spatial degrees of freedom, which enable high-dimensional information encoding~\cite{allen1992,mair2001,boyd2005,Erhard2020}.

In this work, we focus on exploiting these transverse degrees of freedom within the context of Free-Space Optics (FSO)~\cite{Kaushal2017}. Specific solutions of the paraxial Helmholtz equation~\cite{saleh_livro}, such as the Laguerre-Gaussian (LG) and Hermite-Gaussian (HG) mode families, span an infinite-dimensional Hilbert space. These structured optical fields enable high-dimensional encoding by increasing the size of the communication alphabet. However, atmospheric turbulence remains one of the principal physical limitations of FSO communication, introducing random phase distortions that degrade the spatial structure of the transmitted modes and reduce the information carried by the channel~\cite{paterson2005,rodenburg2012,OAM_turbo}. Nevertheless, certain structured beams, such as self-healing Airy beams~\cite{siviloglou2007,gu2010} or recently proposed turbulence eigenmodes~\cite{klug2023robust}, exhibit enhanced robustness against atmospheric distortions.

From an experimental perspective, however, atmospheric turbulence primarily modifies the optical phase, whereas most practical FSO receivers measure only the transverse intensity distribution using standard imaging devices~\cite{Goodman2005}. Consequently, evaluating the degradation of structured beams from intensity-only measurements becomes a fundamental challenge for realistic communication systems, particularly in scenarios where a stable phase reference is unavailable.

Here, we address this problem by experimentally emulating Kolmogorov atmospheric turbulence using a single Spatial Light Modulator, on which both the structured optical fields and the turbulent channels are encoded. The turbulent channels are generated from Kolmogorov phase screens characterized by the Fried parameter~\cite{fried1966statistics}, while computer-generated holograms are employed to encode the structured optical fields~\cite{SLM_artigoOriginal_tecnica}. Using Laguerre-Gaussian, Hermite-Gaussian, and Airy beams as representative structured states, we perform a systematic comparison between four intensity-based diagnostic metrics, Normalized Cross-Correlation (NCC), Strehl Ratio (SR), Beam Width Broadening (BRO), and the Scintillation Index (SC)~\cite{Lewis1995,Andrews2005}, to evaluate their capability to quantify turbulence-induced degradation. Rather than focusing exclusively on the resilience of a particular beam family, our objective is to identify the strengths, limitations, and physical interpretation of each metric, thereby providing practical guidelines for the characterization of structured-light free-space optical links using intensity-only measurements.

The organization of this paper is as follows. Section~\ref{sec:effect_turbulence} reviews the propagation of structured light under atmospheric turbulence and introduces the four diagnostic metrics employed throughout this work. Section~\ref{sec:exp} describes the experimental implementation and presents a comparative analysis of the obtained results. Finally, Section~\ref{sec:conclusion} summarizes the main conclusions and discusses possible future developments.

%%%%%%%%%%%%%%%%%%%%%%%%%%%%%%%%%%%%%%%%%%%%%%%%%%%%%%%%%%%%%%%%
%%%%%%%%%%%%%%%%%%%%%%%%%%%%%%%%%%%%%%%%%%%%%%%%%%%%%%%%%%%%%%%%
%%%%%%%%%%%%%%%%%%%%%%%%%%%%%%%%%%%%%%%%%%%%%%%%%%%%%%%%%%%%%%%%
\section{The effect of turbulence on structure light}\label{sec:effect_turbulence} 

We focus on the impact of atmospheric turbulence on the spatial structure of electromagnetic fields propagating in free space, such as laser beams. In this regime, the magnetic field is fully determined by the electric field, and the explicit time dependence can be disregarded since our focus is exclusively on the spatial properties of the field. Furthermore, because atmospheric turbulence does not alter the polarization state or the magnetic permeability of the medium, the vectorial nature of the electric field can be neglected, allowing us to treat the problem within a scalar framework.

Under these assumptions, the spatial field profile satisfies the Helmholtz equation 
\begin{equation}
    \left(\nabla^{2} + n^{2}k^{2}\right)\psi_\alpha(\boldsymbol{\rho},z) = 0,
\end{equation}
where $\boldsymbol{\rho}$ denotes the transverse coordinates, $n$ is the refraction index, and $z$ is the propagation direction. The index $\alpha$ labels distinct modes. We further restrict our analysis to paraxial solutions for which the field envelope varies slowly along the propagation axis $z$. This class of solutions includes a wide range of structured optical modes, such as Laguerre–Gaussian ($\mathrm{LG}$), Hermite–Gaussian ($\mathrm{HG}$) and Airy ($\mathrm{Ai}$) beams. The explicit forms of the modes considered in this work are provided in the Appendix~\ref{app:modes}.

When a light beam propagates through the atmosphere, it experiences a medium whose refractive index fluctuates randomly in space and time due to variations in temperature and pressure. These fluctuations introduce random phase delays across the transverse profile of the beam. Although the electromagnetic field remains continuous and energy is conserved, the originally smooth wavefront becomes distorted at multiple spatial scales. Large-scale fluctuations produce beam wandering and global tilts, intermediate scales cause wavefront corrugation and beam spreading, and small-scale fluctuations lead to fine interference effects such as speckle formation \zanin{Ref}~\cite{Andrews2005}.

Because the primary action of turbulence is in the optical phase, the effects on intensity arise indirectly through interference. Phase distortions redistribute energy across the transverse plane, generating random intensity modulations known as scintillation. As propagation continues, these distortions accumulate, and the beam progressively loses spatial coherence. Consequently, even if the initial beam is highly structured, the detected intensity pattern becomes increasingly irregular and depends on the particular realization of the turbulent medium \zanin{Ref}~\cite{Andrews2005}.

At sufficiently strong turbulence or long propagation distances, the beam can no longer be meaningfully described as a deterministic object. Instead, its properties must be characterized statistically, through ensemble averages and correlation functions. In this regime, single-shot images carry limited information, and reproducible features emerge only through quantities that quantify similarity, coherence, or spatial correlation between different realizations or between distorted and reference fields.

Experimentally, the directly accessible quantity is the transverse intensity distribution, $I(\boldsymbol{\rho},z) = \left|\psi(\boldsymbol{\rho},z)\right|^2$, which can be measured with a CCD camera. The main idea to quantify the effect of atmospheric turbulence on the spatial structure of the beam is to compare the measured intensity profile after propagation through turbulence with a reference intensity profile obtained under identical preparation and propagation conditions, but in the absence of turbulence. These two intensity distributions are denoted by $I_T(\boldsymbol{\rho},z)$ and $I_0(\boldsymbol{\rho},z)$, respectively.

We discuss four commonly used intensity-based measures:
the normalized cross-correlation (NCC)~\cite{Lewis1995},
the Strehl ratio, beam-width broadening, and the
scintillation index~\cite{Andrews2005}. For each quantity, we comment on its physical meaning and relevance for structured beams such as Laguerre-Gaussian, Hermite-Gaussian, and Airy modes.

In the Kolmogorov picture of atmospheric turbulence \zanin{Ref} ~\cite{Tatarski1971}, different spatial scales of fluctuations in the refractive index produce distinct physical effects on a propagating beam. The largest eddies, characterized by the outer scale, primarily induce beam wander through large-scale tip–tilt distortions of the wavefront. Intermediate scales generate wavefront distortions that reduce phase coherence across the beam profile, while the smallest eddies, near the inner scale, give rise to rapid phase variations that manifest as speckle and intensity granularity. Within this framework, the normalized cross-correlation (NCC) is sensitive to modal distortion arising from fluctuations across all these scales, as it probes the preservation of the overall spatial structure of the beam. The Strehl ratio (SR) predominantly reflects the accumulated phase variance caused mainly by intermediate-scale distortions that degrade coherent interference. Beam broadening, which is measured by the broadering parameter (BRO), is chiefly associated with angular scattering and energy redistribution in the transverse plane, often linked to cumulative phase gradients across multiple scales. Finally, the scintillation index (SC) is especially sensitive to small-scale turbulence, as it quantifies the intensity fluctuations produced by fine-scale interference effects and speckle formation~\cite{Andrews2005,Tatarski2016,Tatarski1971,Ishimaru1978,Andrews2001,Mandel1995}.
 
%%%%%%%%%%%%%%%%%%%%%%%%%%%%%%%%%%%%%%%%%%%%%%%%%%%%%%%%%%%%%%%
\subsection{Normalized cross-correlation}

A natural measure of similarity between two transverse intensity patterns is provided by their spatial overlap. However, a simple overlap integral depends on the absolute intensity scale and detector gain and therefore does not constitute a robust similarity measure. To remove this dependence on the random turbulence realization,
we employ the normalized cross-correlation (NCC)~\cite{rodenburg2012,Schulze2013}.
\begin{equation}
\mathrm{NCC}
=
\frac{
\displaystyle
\int \dd^2\boldsymbol{\rho}\;
I_0\,
I_T
}{
\sqrt{
\displaystyle
\int \dd^2\boldsymbol{\rho}\;
I_0^2
\;
\int \dd^2\boldsymbol{\rho}\;
I_T^2
}
},
\label{eq:ncc_def}
\end{equation}
where we removed the coordinate dependence of the intensities to simplify notation.

By construction, the NCC satisfies $0 \le \mathrm{NCC} \le 1$. A value $\mathrm{NCC} = 1$ corresponds to identical spatial intensity profiles, while values close to zero indicate negligible spatial similarity. Physically, the NCC quantifies how much of the original spatial organization of the structured beam survives after propagation through the turbulent medium. In other words, it measures pattern similarity. As turbulence introduces random phase distortions that redistribute optical energy across the transverse plane, the overlap with the reference intensity pattern tends to decrease, leading to a reduction of the NCC.

%For $\mathrm{LG}$ beam, turbulence causes vortex core distortion, thus lowering $\mathrm{NCC}$. For $\mathrm{HG}$ beams, the nodal lines breaking tends to lowers NCC. In the case of Airy beam, the main lobe deforms, thus lowering NCC. \zela{Estas afirmações estão soltas. Há referências? Isso é demonstrado neste trabalho?}
%%%%%%%%%%%%%%%%%%%%%%%%%%%%%%%%%%%%%%%%%%%%%%%%%%%%%%%%%%%%%%%
\subsection{Strehl Ratio}

The Strehl ratio ($\mathrm{SR}$) quantifies the reduction in peak intensity due to turbulence. It is defined as
\begin{equation}\label{eq:SR}
\mathrm{SR} = 
\frac{\max\left[I_{T}\right]}{\max\left[I_{0}\right]},
\end{equation}
Physically, the Strehl ratio measures the loss of phase coherence across the wavefront. The peak intensity is maximal only if all parts of the wavefront interfere constructively. Since turbulence introduces random phase across the wavefront, constructive interference tends to decrease.

%For Gaussian modes, $\mathrm{SR}$ directly measures wavefront degradation. For $\mathrm{LG}$ modes with azimuthal index $\ell$, turbulence reduces the darkness on the axis and fills the vortex core, making $\mathrm{SR}$ sensitive to vortex degradation. For the $\mathrm{HG}$ modes, $\mathrm{SR}$ captures the distortion of the nodal structure. For Airy beams, whose peak follows a curved trajectory, $\mathrm{SR}$ is sensitive to distortion of the main lobe, as its coherence tends to decrease. \zela{Referências??}

The Strehl ratio only probes the \emph{maximum} intensity. It does not capture changes in the modal structure or energy redistribution across the transverse plane. For multi-lobed modes ($\mathrm{HG}$, $\mathrm{LG}$ with $p>0$, Airy beams), SR alone is insufficient to characterize structural degradation.

%%%%%%%%%%%%%%%%%%%%%%%%%%%%%%%%%%%%%%%%%%%%%%%%%%%%%%%%%%%%%%%
\subsection{Beam width broadening}

The width of the beam measures the transverse spatial spread of the intensity. It is commonly defined through the second central moment,
\begin{equation}\label{eq:Second_moment}
w^2 = 
\frac{\int |\boldsymbol{\rho}-\boldsymbol{\rho}_c|^2 
I(\boldsymbol{\rho})\, d^2\boldsymbol{\rho}}
{\int I(\boldsymbol{\rho})\, d^2\boldsymbol{\rho}},
\end{equation}
where $\boldsymbol{\rho}_c$ is the centroid,
\begin{equation}
\boldsymbol{\rho}_c =
\frac{\int \boldsymbol{\rho} I(\boldsymbol{\rho})\, d^2\boldsymbol{\rho}}
{\int I(\boldsymbol{\rho})\, d^2\boldsymbol{\rho}}.
\end{equation}
Turbulence generally increases $w$, reflecting beam spreading and
loss of spatial coherence. In other words, it is a measure of how much transverse spatial diffusion occurred. The broadening parameter is defined as
\begin{equation}\label{eq:BRO}
    \mathrm{BRO} = \frac{w^{2}_{T}}{w^{2}_0}.
\end{equation} 

%For the $\mathrm{LG}$ modes, turbulence induces coupling between different radial and azimuthal indices, increasing the effective beam radius. Higher radial modes get excited. For modes $\mathrm{HG}$, turbulence smooths the nodal lines and increases the second moments. Thus, the nodal regions blur. For Airy beams, turbulence suppresses self-acceleration and broadens the main lobe. \zela{Referências??}

The beam width is insensitive to fine structural details.
Two beams with very different modal content may share the same
second moment. Thus, width alone cannot quantify the modal purity.

%%%%%%%%%%%%%%%%%%%%%%%%%%%%%%%%%%%%%%%%%%%%%%%%%%%%%%%%%%%%%%%
\subsection{Scintillation Index}

The scintillation index measures intensity fluctuations and is defined as
\begin{equation}\label{eq:scintilation_index}
\sigma^2 =
\frac{\langle I^2 \rangle - \langle I \rangle^2}
{\langle I \rangle^2},
\end{equation}
where $\langle \cdot \rangle$ denotes spatial averaging
over the CCD camera. From this we can define the scintillation parameter as
\begin{equation}\label{eq:SC}
    \mathrm{SC} = \frac{\sigma_T^2}{\sigma_0^2}.
\end{equation}

%Scintillation quantifies the formation of speckles and the amplitude fluctuations induced by refractive index inhomogeneities. In other words, scintillation measures intensity randomness. For $\mathrm{LG}$ modes with higher $|\ell|$ often exhibit enhanced scintillation due to increased phase sensitivity. Turbulence leads to the vortex core becoming unstable. $\mathrm{HG}$ modes develop speckled interference patterns along nodal directions. Airy beams may show fragmentation of the main lobe and increased intensity variance. \zela{Referências??}

Scintillation measures statistical intensity fluctuations but does not directly quantify phase distortion or modal coupling. It is therefore complementary to structural metrics such as normalized cross-correlation or modal decomposition.

%%%%%%%%%%%%%%%%%%%%%%%%%%%%%%%%%%%%%%%%%%%%%%%%%%%%%%%%%%%%%%%%
%%%%%%%%%%%%%%%%%%%%%%%%%%%%%%%%%%%%%%%%%%%%%%%%%%%%%%%%%%%%%%%%
%%%%%%%%%%%%%%%%%%%%%%%%%%%%%%%%%%%%%%%%%%%%%%%%%%%%%%%%%%%%%%%%
\section{Experiment}\label{sec:exp}

\begin{figure}[htbp]
    \centering
    \includegraphics[width=1\linewidth]{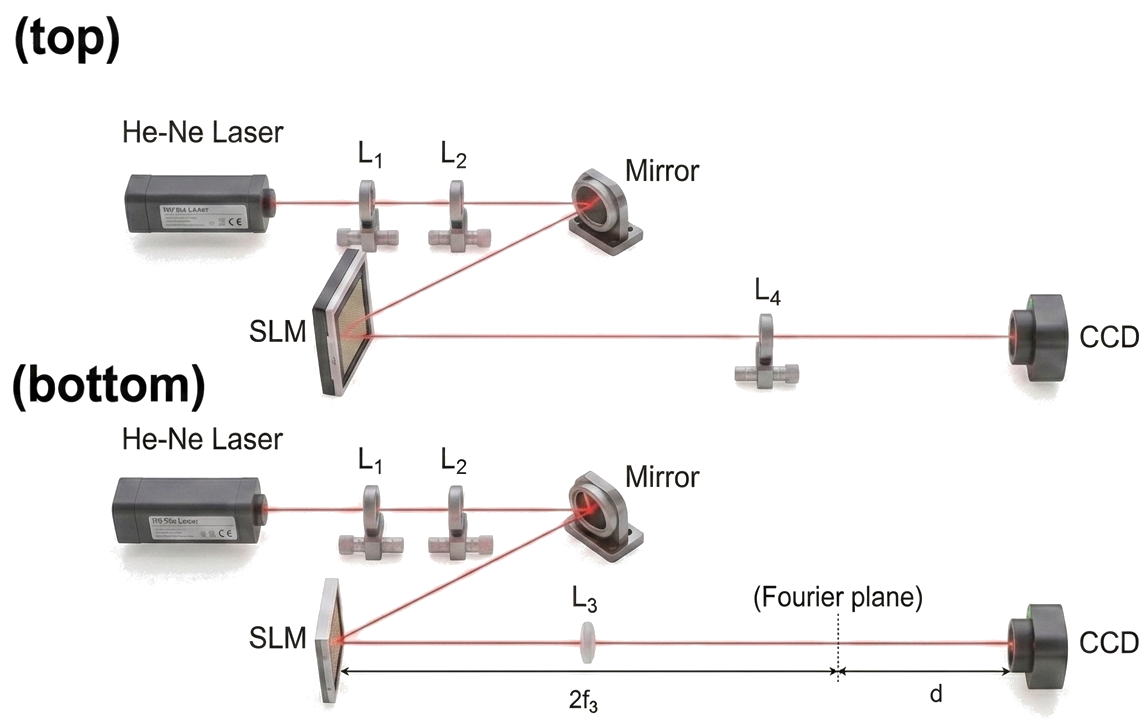}
\caption{\justifying . A He-Ne laser beam ($\lambda = 633$~nm) is expanded via a telescope consisting of lenses $L_1$ and $L_2$, and then directed onto a spatial light modulator (SLM). The SLM encodes the hologram masks required for both structured light generation and atmospheric turbulence emulation. \textbf{(top)} Configuration where lens $L_4$ performs spatial filtering of undesired diffraction orders to isolate HG and LG modes. \textbf{(bottom)} Configuration where the Fourier transform of the Airy beam is implemented [Eq.~\eqref{app:eq:Ai}], followed by lens $L_3$ to perform the inverse Fourier transform. The Fourier plane defines the zero-reference position ($z=0$) for the CCD displacements during measurements.}
    \label{fig:setup}
\end{figure}

The experimental setup consists of a laser source ($\lambda=633$~nm) expanded by a telescope system (composed of lenses $f_1$ and $f_2$ to increase its waist) and directed toward a reflective Spatial Light Modulator (SLM). After the SLM, a lens $f_3$ performs a Fourier transform to allow for the spatial filtering of undesired diffraction orders. For Airy beams, the mode is generated at the Fourier plane, and $f_3$ effectively performs the inverse transform, creating the beam at a distance of $2f_3$ from the SLM. For the LG and HG modes, the beam is spatially filtered at the Fourier plane, and an additional lens $f_4$ is used to perform the inverse Fourier transform. In all cases, the beam is finally directed onto a CCD camera to record the intensity distributions. The experimental configuration is illustrated in Fig.~\ref{fig:setup}.

The SLM serves as the core component of this apparatus. By employing a complex modulation technique~\cite{SLM_artigoOriginal_tecnica}, it generates the various beam profiles investigated in this study. Furthermore, the SLM enables the simulation of atmospheric turbulence based on Kolmogorov statistics~\cite{fried1966statistics,Andrews2005}. This is achieved by imposing a secondary phase mask that reproduces the random phase fluctuations characteristic of a turbulent medium. The effective hologram results from the superposition of both the beam-shaping and turbulence masks~\cite{SLM_artigoOriginal_tecnica}. We implemented Fried parameters ($r_0$) ranging from 10.67~cm to 4.64~cm. This corresponds to a moderate-to-strong turbulent medium with a refractive index structure constant of $C_n^2 = 1 \times 10^{-15}$~m$^{-2/3}$~\cite{Andrews2005}. Because the Fried parameter is inversely proportional to the propagation distance, a smaller $r_0$ represents a longer propagation path through the turbulence, assuming a constant $C_n^2$~\cite{Andrews2005}. Therefore, our selected parameters simulate a beam propagating through a turbulent medium over distances ranging from $L=1$~km to $L=4$~km, as detailed in Table~\ref{tab:parametros_r0}.

%%%%%%%%%%%%%%%%%%%%%%%%%%%%%%%%%%%%%%%%%%%%%%%%%%%%%%%%
%%%%%%%%%%%%%%%%%%%%%%%%%%%%%%%%%%%%%%%%%%%%%%%%%%%%%%%

To isolate the effects of turbulence, two distinct measurements are performed for a fixed optical path of length $L$:
\begin{enumerate}
    \item A \textbf{reference measurement}, in which the beam propagates through free space (simulated by applying only the beam-shaping mask to the SLM). The resulting transverse intensity profile, denoted as $I_0$, is recorded by the CCD camera.
    \item A \textbf{turbulent measurement}, in which the same prepared beam propagates through a simulated turbulent medium (achieved by activating the combined holographic mask). The perturbed intensity profile, denoted as $I_T$, is recorded at the same axial position.
\end{enumerate}

All experimental parameters, as detailed in Table~\ref{tab:parametros}, are maintained identical between the two measurements, including optical alignment, propagation distance, imaging system, and camera settings. Consequently, any deviation between $I_0$ and $I_T$ is attributed exclusively to the effects of the simulated turbulence.

%%%%%%%%%%%%%%%%%%%%%%%%%%%%%%%%%%%%%%%%%%%%%%%%%%%%%%%%
\begin{table}[htbp]
    \centering
    \caption{\justifying Fried parameter ($r_0 = \left( 0.423 \, k^2 \, C_n^2 \, L \right)^{-3/5})$ as a function of the simulated propagation distance $L$, considering $C_n^2 = 1 \times 10^{-15}$~m$^{-2/3}$.}
    \label{tab:parametros_r0}
    \begin{tabular}{cc}
        \toprule
        \textbf{$L$ (m)} & \textbf{$r_0$ (cm)} \\ 
        \midrule
        0 (no turbulence) & $\rightarrow \infty$ \\ 
        1000 & 10.67 \\ 
        2000 & 7.04 \\ 
        3000 & 5.52 \\ 
        4000 & 4.64 \\ 
        \bottomrule
    \end{tabular}
\end{table}
%%%%%%%%%%%%%%%%%%%%%%%%%%%%%%%%%%%%%%%%%%%%%%%%%%%%%%%%

%%%%%%%%%%%%%%%%%%%%%%%%%%%%%%%%%%%%%%%%%%%%%%%%%%%%%%%%
\begin{table}[htbp]
    \centering
    \caption{\justifying Experimental parameters for the generated optical beams. For LG and HG modes, $w_0$ denotes the beam waist radius in Eqs.~\eqref{app:eq:LG} and \eqref{app:eq:HG}, respectively. For the Airy beams, $a_x$ is the exponential decay parameter and $x_0$ is the transverse scale parameter in Eq.~\eqref{app:eq:Ai}.}
    \label{tab:parametros}
    \begin{tabular}{lcc}
        \toprule
        \textbf{Beam Type} &  &  \\
        \midrule
        LG   & $w_0 = 0.3~\text{mm}$   & -- \\ 
        HG   & $w_0 = 0.3~\text{mm}$   & -- \\ 
        Airy & $a_x = 0.1$             & $x_0 = 0.19~\mu\text{m}$ \\ 
        \bottomrule
    \end{tabular}
\end{table}
%%%%%%%%%%%%%%%%%%%%%%%%%%%%%%%%%%%%%%%%%%%%%%%%%%%%%%%

For each experimental configuration, we record 10 reference measurements ($I_0$) and 10 turbulent measurements ($I_T$). The repeated reference measurements are taken to ensure the stability of the unperturbed beam and to verify that the system does not drift over time. 

For the simulated turbulent medium, the phase masks are generated randomly; thus, an infinite number of unique mask realizations exist for any given Fried parameter ($r_0$). Multiple realizations are necessary because any single phase mask contains localized regions of stronger and weaker phase fluctuations. For example, if a region of strong phase fluctuation happens to overlap with the spatial profile of the generated mode on the SLM, it will degrade the beam significantly more than a weaker region would, even though the overall global turbulence strength remains constant. This statistical approach is essential to accurately emulate real atmospheric turbulence, which is highly dynamic and continuously evolving. 

\begin{figure}[t]
    \centering
    \includegraphics[width=\columnwidth]{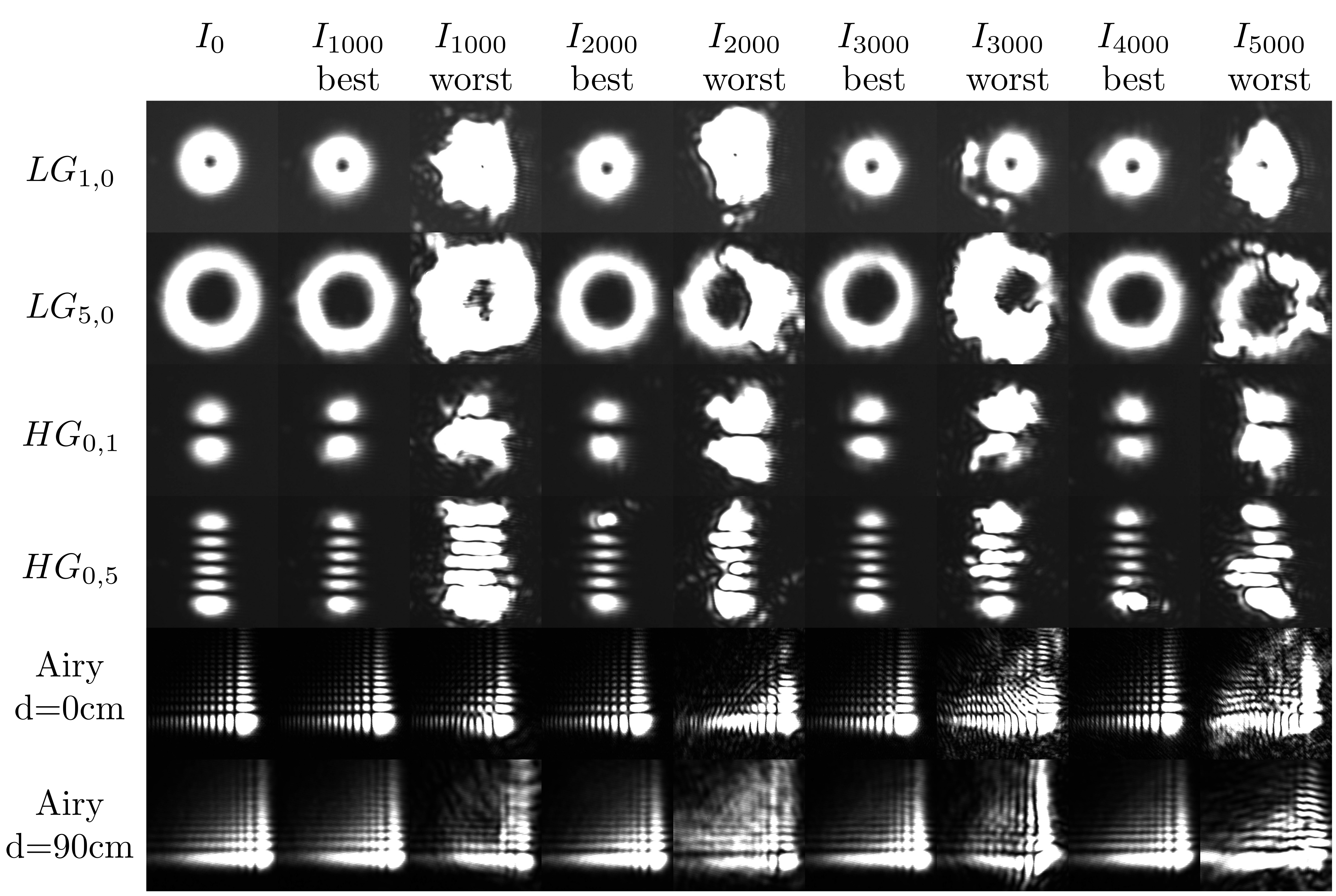}
    \caption{\justifying Measurements of the Wavefront intensity profiles of structured beams under the influence of simulated atmospheric turbulence. Rows from top to bottom correspond to lower-order ($LG_{1,0}$, $HG_{0,1}$) and higher-order ($LG_{5,0}$, $HG_{0,5}$) modes, followed by Airy beams recorded at distances of $d=0$~cm and $d=90$~cm from the Fourier plane. The first column ($I_0$) displays the turbulence-free reference profiles. Subsequent columns ($I_L$) show the statistical variation for simulated propagation distances from $L=1000$~km to $L=5000$~km, explicitly contrasting the least distorted (best) and most distorted (worst) realizations within each experimental ensemble.}
    \label{fig:wavefront}
\end{figure}

By acquiring 10 independent measurements for both the reference intensity $I_0$ and the turbulent realizations $I_T$, we construct a comprehensive statistical ensemble. Comparing each reference image against every turbulent realization yields 100 unique cross-comparison pairs, ensuring a statistically significant analysis. This approach is necessary due to the stochastic nature of atmospheric turbulence; indeed, considerably different wavefront distortions can arise from phase masks characterized by the same Fried parameter. Because random phase fluctuations may exhibit stronger modulation in localized regions of the hologram, the impact on the beam depends heavily on whether these severe turbulence features overlap with the intensity profile of the structured mode. 

Figure~\ref{fig:wavefront} illustrates these measurements for lower-order modes ($LG_{1,0}$ and $HG_{0,1}$), higher-order modes ($LG_{5,0}$ and $HG_{0,5}$), and Airy beams at the minimum ($d=0$~cm) and maximum ($d=90$~cm) propagation distances from the Fourier plane. The results are organized as follows: $I_0$ represents the wavefront without turbulence, while $I_{L}$ denotes the profiles affected by a simulated propagation distance of $L$~km. To illustrate the wavefront degradation statistics, we present the least affected (best) and most affected (worst) realizations from each experimental set.

\begin{figure}[htbp]
    \centering
    \includegraphics[width=0.98\columnwidth]{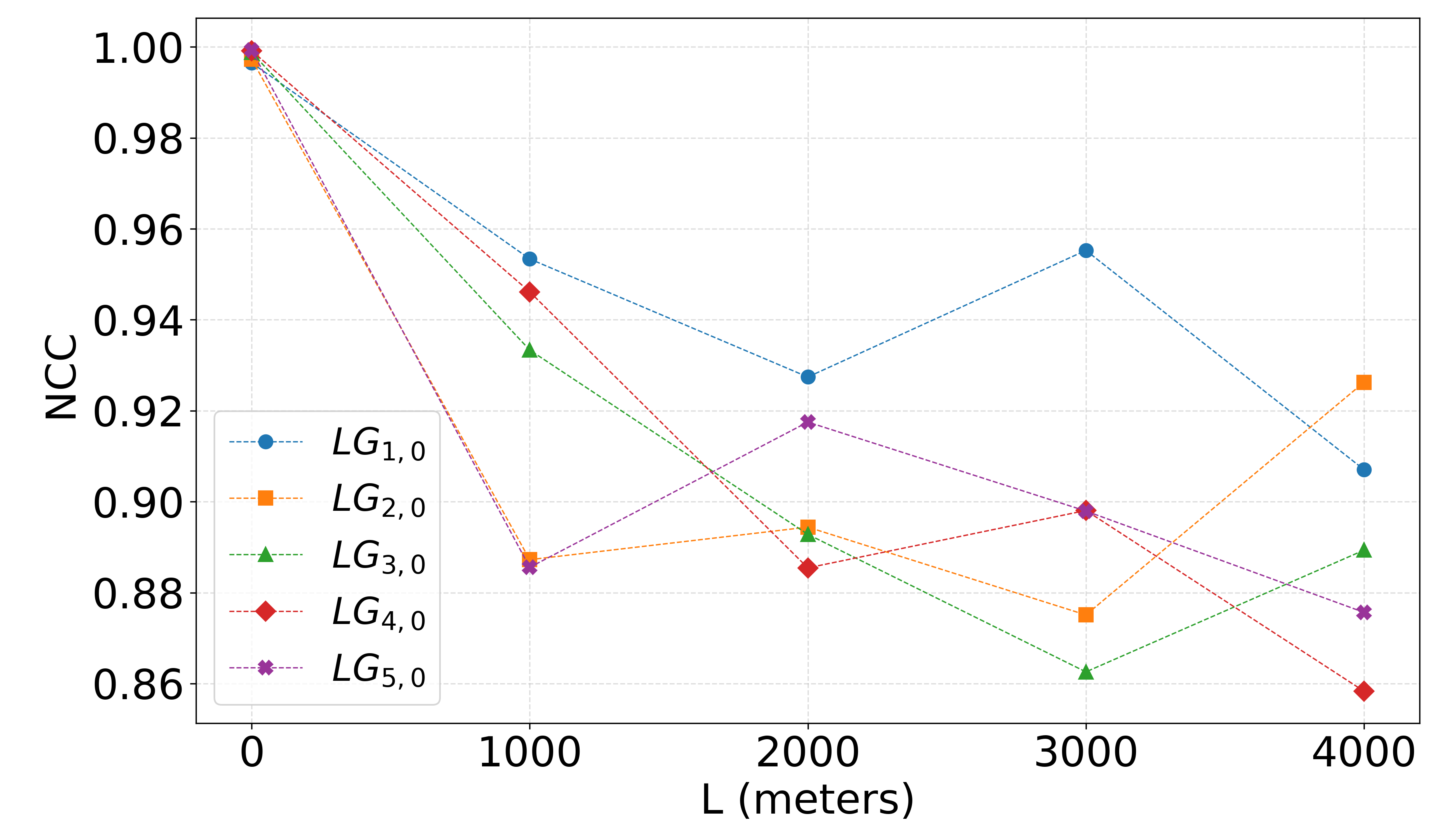}
    \vspace{0.2em}
    \includegraphics[width=0.98\columnwidth]{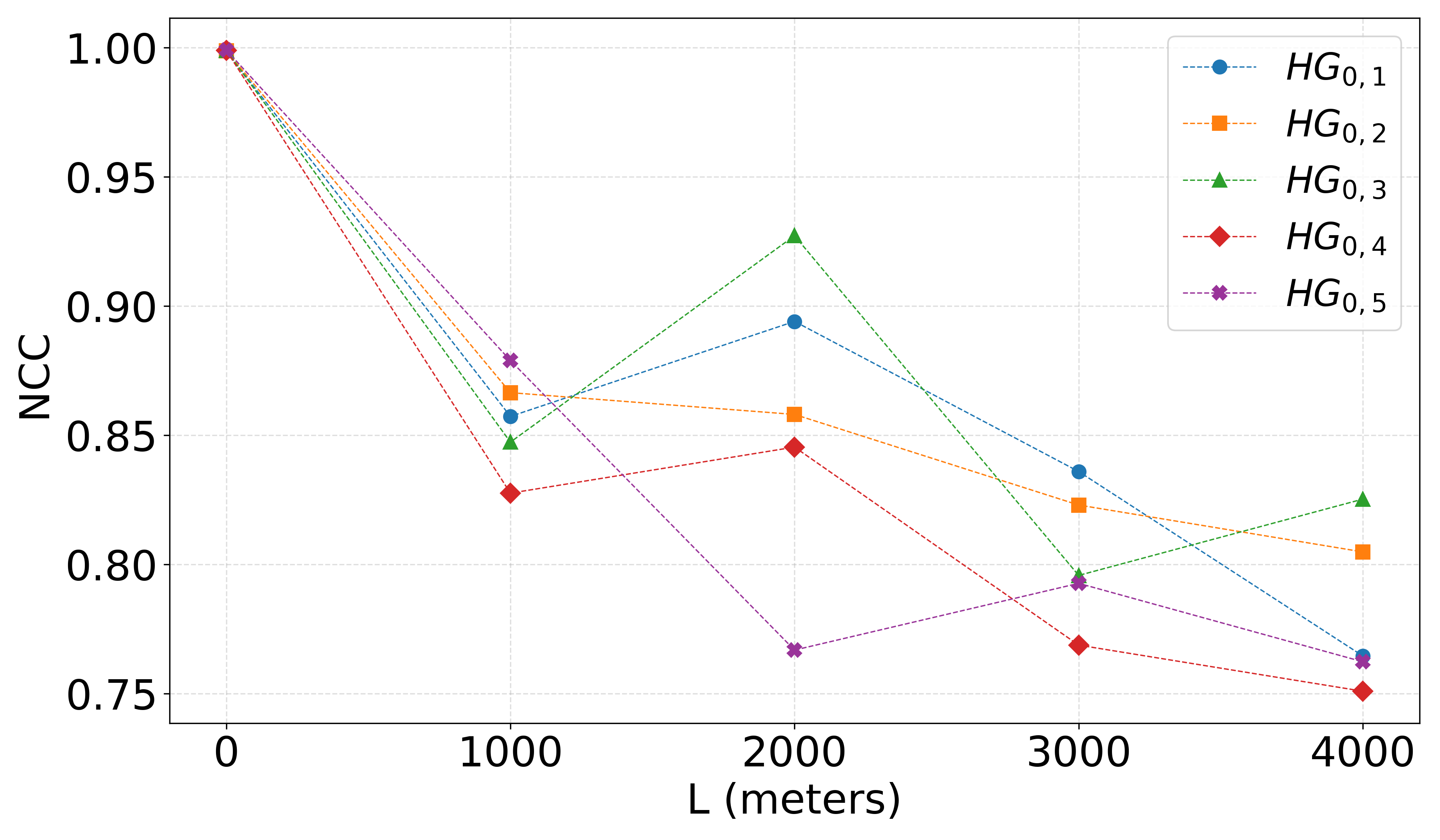}
    \vspace{0.2em}
    \includegraphics[width=0.98\columnwidth]{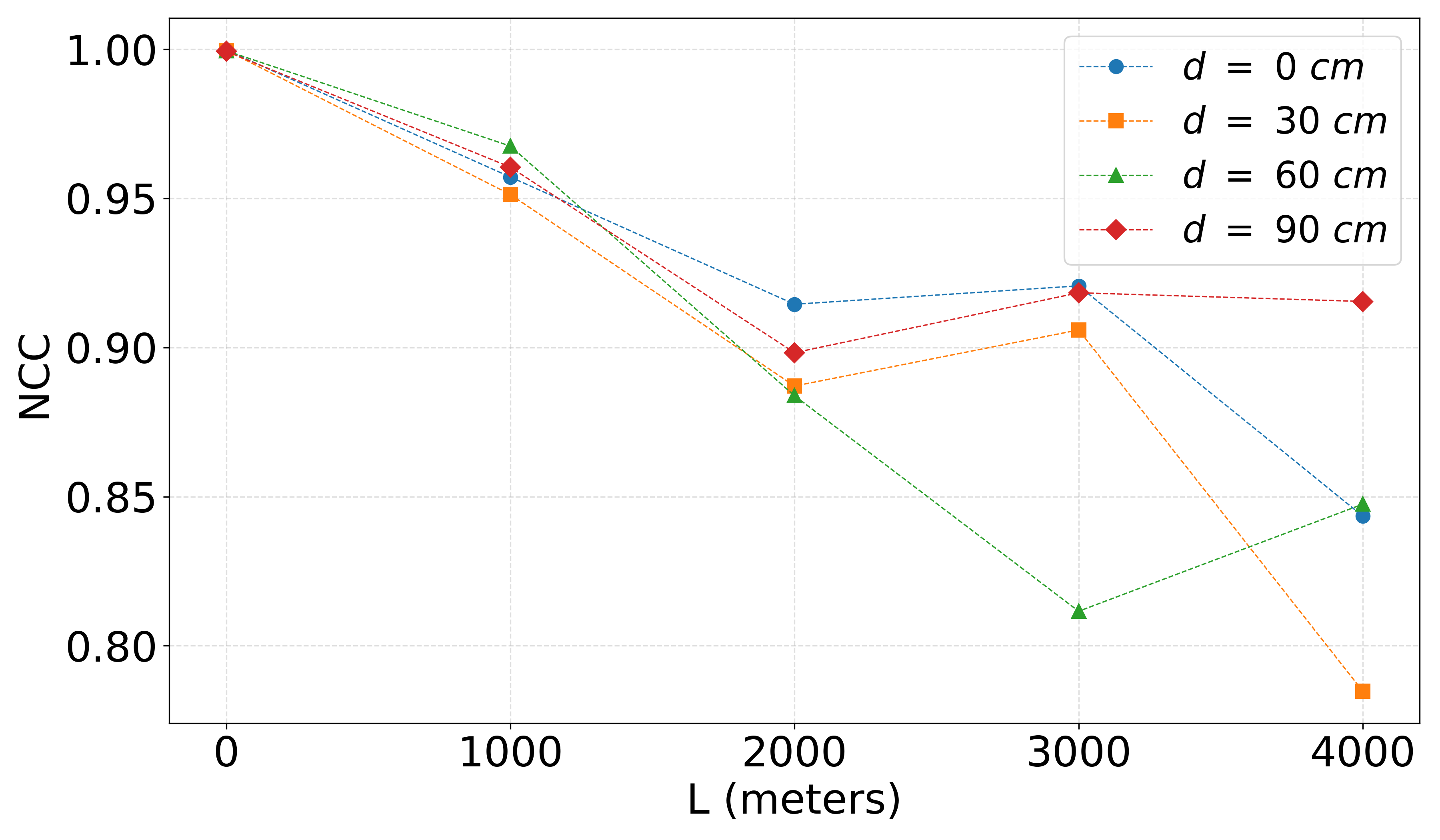}
    
\caption{\justifying Normalized Cross-Correlation (NCC) results as a function of the simulated turbulence propagation distance (ranging from $0\text{ m}$ to $4000\text{ m}$) for the three families of structured light beams: 
\textbf{(top)} Laguerre-Gaussian (LG) modes ($\ell \in \{1, \dots, 5\}$), illustrating a sharp morphological degradation and loss of azimuthal symmetry under mild turbulence; 
\textbf{(middle)} Hermite-Gaussian (HG) modes ($m=0, n \in \{1, \dots, 5\}$), which provide a highly sensitive global measure of structural decay due to the progressive blurring and erasure of their characteristic rectangular nodal lines; and 
\textbf{(bottom)} Airy beams, demonstrating superior structural resilience and morphological stability across long propagation distances, sustained by their non-diffracting nature and self-healing mechanism.}
\label{fig:NCC}
\end{figure}

The experimental results are organized by every method, with each figure displaying the performance of the Laguerre-Gaussian (LG), Hermite-Gaussian (HG), and Airy modes under simulated atmospheric turbulence. Specifically, Fig.~\ref{fig:NCC} presents the results for the Normalized Cross-Correlation (NCC), Fig.~\ref{fig:SR} evaluates the Strehl Ratio (SR), Fig.~\ref{fig:BRO} delineates the Beam Width Broadening (BRO) factor, and Fig.~\ref{fig:SC} quantifies the Scintillation Index (SC). Within each figure, the data are structured vertically, moving from the top panel for LG modes, through the middle panel for HG modes, to the bottom panel dedicated to Airy beams, spanning a simulated propagation distance from $0\text{ m}$ to $4000\text{ m}$.

In the next subsection, we discuss the results method by method, comparing all modes to evaluate what each method reveals and how they relate to the experimental data. In all figures, the first data point compares every $I_0$ measurement with one another; this is done to guarantee that the reference beams are consistent. Consequently, all parameters are approximately equal to one when comparing measurements that are highly similar.

%%%%%%%%%%%%%%%%%%%%%%%%%%%%%%%%%%%%%%%%%%%%%%%%%%%%%%%%%%%%%%%%
%%%%%%%%%%%%%%%%%%%%%%%%%%%%%%%%%%%%%%%%%%%%%%%%%%%%%%%%%%%%%%%%
%%%%%%%%%%%%%%%%%%%%%%%%%%%%%%%%%%%%%%%%%%%%%%%%%%%%%%%%%%%%%%%%
\subsection{NCC}

The usefulness of the normalized cross-correlation (NCC) as a diagnostic tool for structured light propagating through atmospheric turbulence strongly depends on how the defining features of a given mode family are encoded in the transverse intensity distribution. Since the NCC only compares intensity patterns, it provides a measure of structural similarity rather than a direct probe of phase coherence or modal purity. Its interpretation must therefore be tailored to the specific class of structured beam under consideration.

\subsubsection{Laguerre–Gaussian beams}

For Laguerre-Gaussian (LG) beams, atmospheric turbulence primarily distorts the azimuthal phase structure responsible for Orbital Angular Momentum (OAM). This process leads to significant mode coupling and a redistribution of energy across different radial and azimuthal indices. While these phase distortions eventually manifest as visible structural changes—specifically through fragmentation and the loss of cylindrical symmetry in the ring-shaped intensity profile—a severe degradation of the OAM content often precedes any major deformation of the intensity distribution. Consequently, while the Normalized Cross-Correlation (NCC) decreases reliably as turbulence strength or propagation distance increases, it may provide an overly optimistic estimate of the underlying modal stability. In this sense, the NCC effectively captures the degradation of spatial intensity organization but fails to fully reflect the loss of phase-defined topological properties.

Figure \ref{fig:NCC} (top) presents the NCC results for LG modes with topological charges $\ell$ ranging from 1 to 5, maintaining a fixed radial index $p=0$. As expected, the phase distribution undergoes stochastic degradation, leading to energy redistribution and subsequent crosstalk between different OAM states. Furthermore, a general trend is observed: as the propagation distance ($L$) increases, the modal integrity diminishes, resulting in a consistent decline in NCC values. However, this decay is not strictly monotonic for individual realizations due to the stochastic nature of the atmospheric phase masks; each random realization affects the azimuthal wavefront uniquely. Notably, our results do not show a significant dependence of the degradation rate on the value of $\ell$, with all investigated modes being affected in a statistically similar manner. In these simulations, the NCC values range from approximately $0.96$ in the weak-turbulence regime to approximately $0.86$ where the modes begin to exhibit substantial structural degradation.

\subsubsection{Hermite–Gaussian beams}

Figure \ref{fig:NCC} (middle) presents the results for Hermite–Gaussian beams, which by contrast, are particularly well suited to characterization via NCC. Their defining features—rectangular symmetry and the presence of nodal lines—are directly encoded in the transverse intensity distribution. Turbulence-induced phase distortions readily disrupt the delicate interference conditions that give rise to these nodal structures, causing them to shift, blur, or disappear entirely. These changes are immediately visible in the intensity pattern, leading to a pronounced and monotonic reduction of the NCC as turbulence strength increases. For Hermite–Gaussian modes, intensity-based correlations therefore provide a sensitive and reliable global measure of structural degradation.

\subsubsection{Airy Beams}

Figure \ref{fig:NCC} (bottom), presents Airy beams results. They occupy an intermediate position, exhibiting both sensitivity and robustness in different aspects of their structure. Their asymmetric intensity profile and self-accelerating behavior arise from a broad spatial spectrum and a distributed phase structure. Under turbulence, the side lobes of the Airy beam are typically distorted first, while the dominant main lobe often remains identifiable over longer propagation distances. This self-healing behavior results in a slower decay of the NCC compared to Gaussian-based modes, particularly at moderate turbulence strengths. In this case, the NCC effectively quantifies the progressive loss of fine structure and the eventual breakdown of the accelerating profile, while also highlighting the relative resilience of Airy beams to phase perturbations.

Taken together, these considerations indicate that the NCC serves as a meaningful but inherently coarse-grained metric for assessing the impact of turbulence on structured light. It is most informative when the defining characteristics of a mode are strongly encoded in the intensity distribution, as in Hermite–Gaussian beams, and less definitive when essential information resides primarily in the optical phase, as in Laguerre–Gaussian modes. For Airy beams, the NCC provides insight into structural robustness and self-healing but does not capture subtle phase distortions of the main lobe. Consequently, while the NCC offers a convenient and experimentally accessible measure of turbulence-induced degradation, it should be interpreted as a global indicator of spatial similarity rather than a comprehensive characterization of the underlying field dynamics.

\subsection{Strehl Ratio (SR)}

The Strehl Ratio (SR), as defined in Equation \ref{eq:SR}, provides a quantitative measure of the peak intensity degradation due to atmospheric turbulence. This metric is intrinsically linked to energy scattering; in the weak turbulence regime, where the modal structure remains nearly unperturbed, $SR \approx 1$. As turbulence strength increases, the beam's energy is scattered and spatially redistributed, causing the peak intensity to diminish such that $SR \rightarrow 0$.

However, atmospheric turbulence can act as a stochastic system of micro-lenses, creating localized "hotspots" or scintillation spikes. To prevent these random intensity peaks from distorting the comparative analysis, we normalize both the turbulent and reference images by their total integrated intensity ($\sum I$). This normalization ensures that the SR reflects the structural integrity of the mode rather than transient power fluctuations or localized scintillation effects.

\begin{figure}[htbp]
    \centering
    \includegraphics[width=0.98\columnwidth]{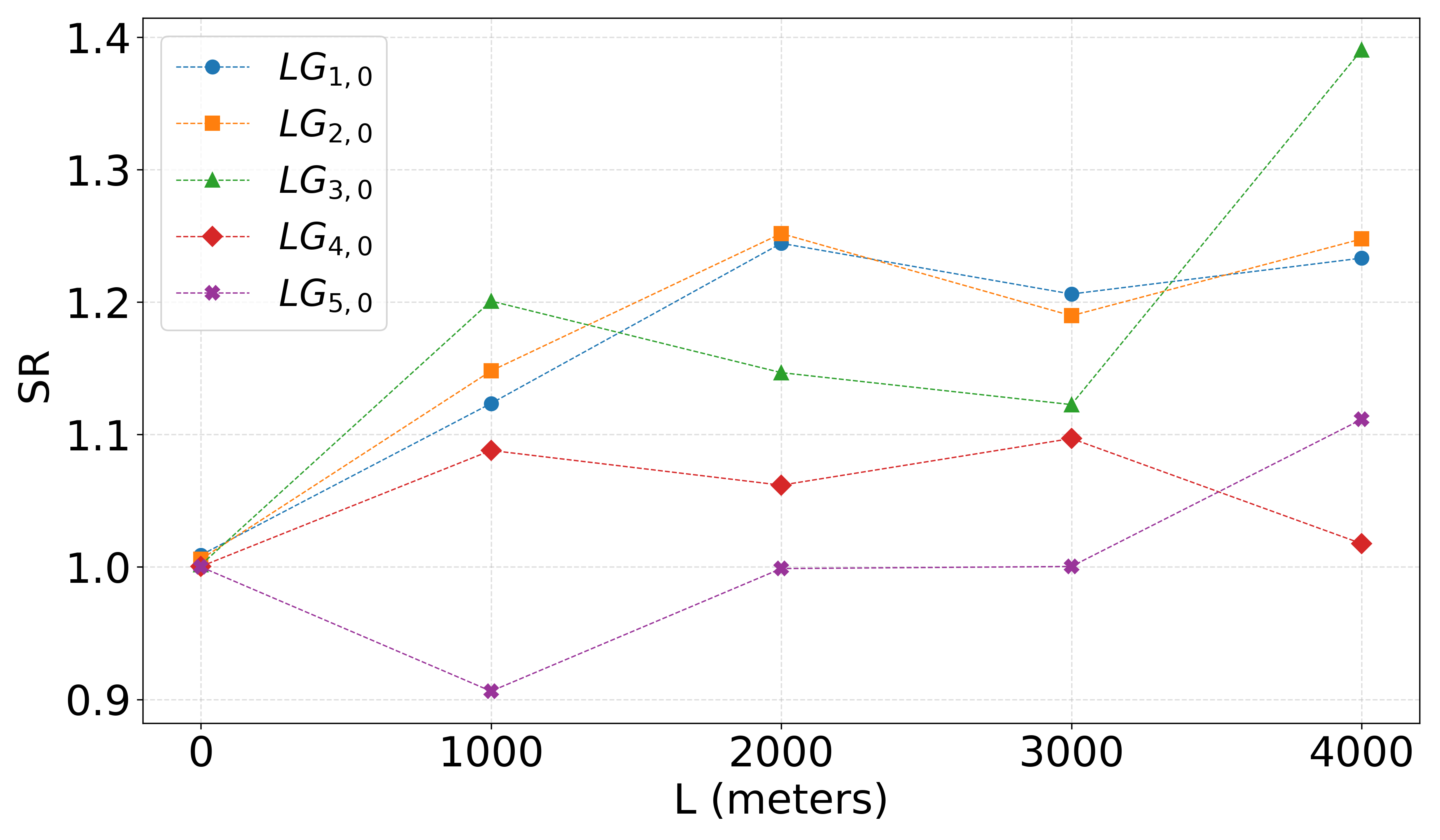}
    \vspace{0.2em}
    \includegraphics[width=0.98\columnwidth]{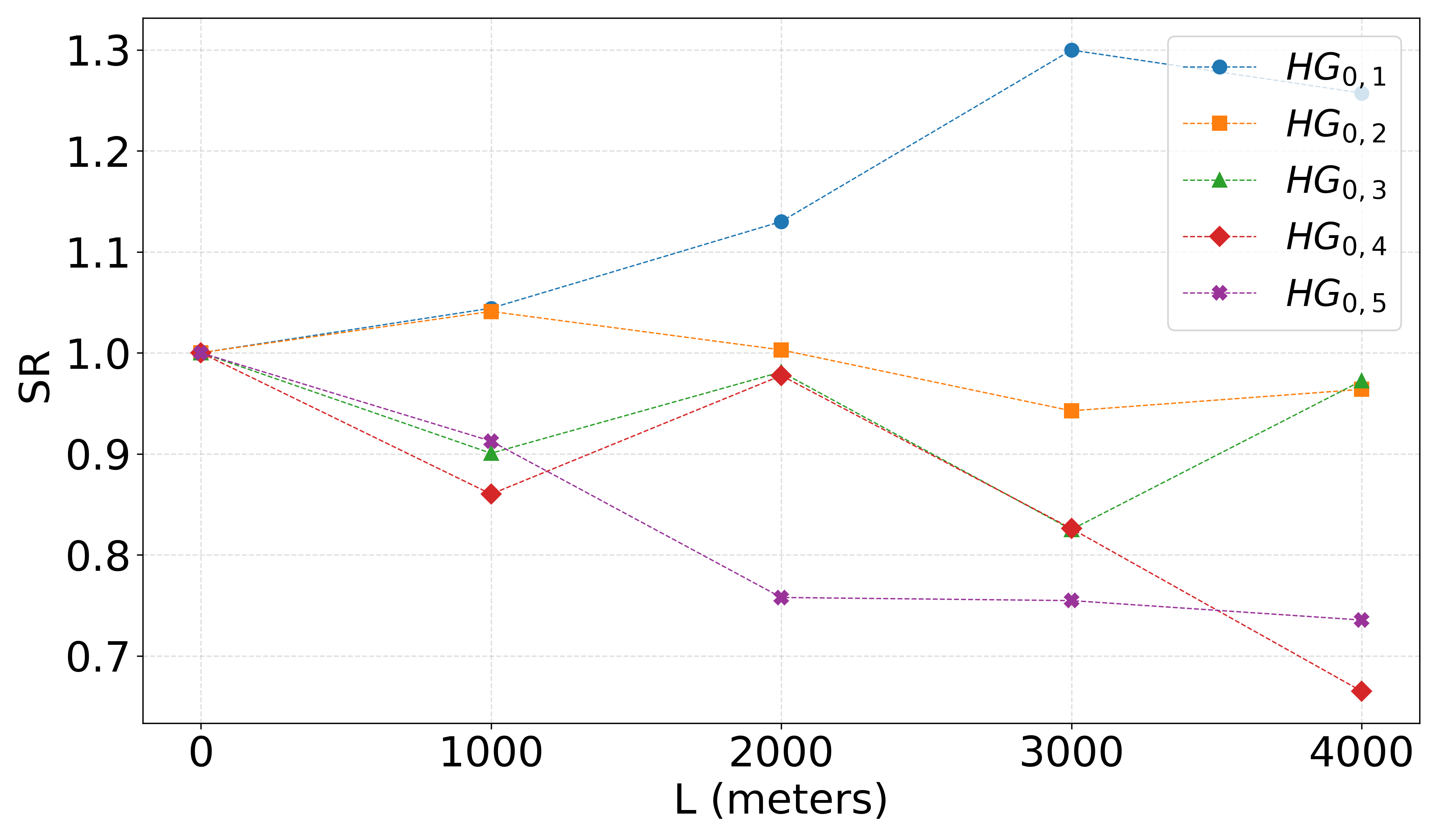}
    \vspace{0.2em}
    \includegraphics[width=0.98\columnwidth]{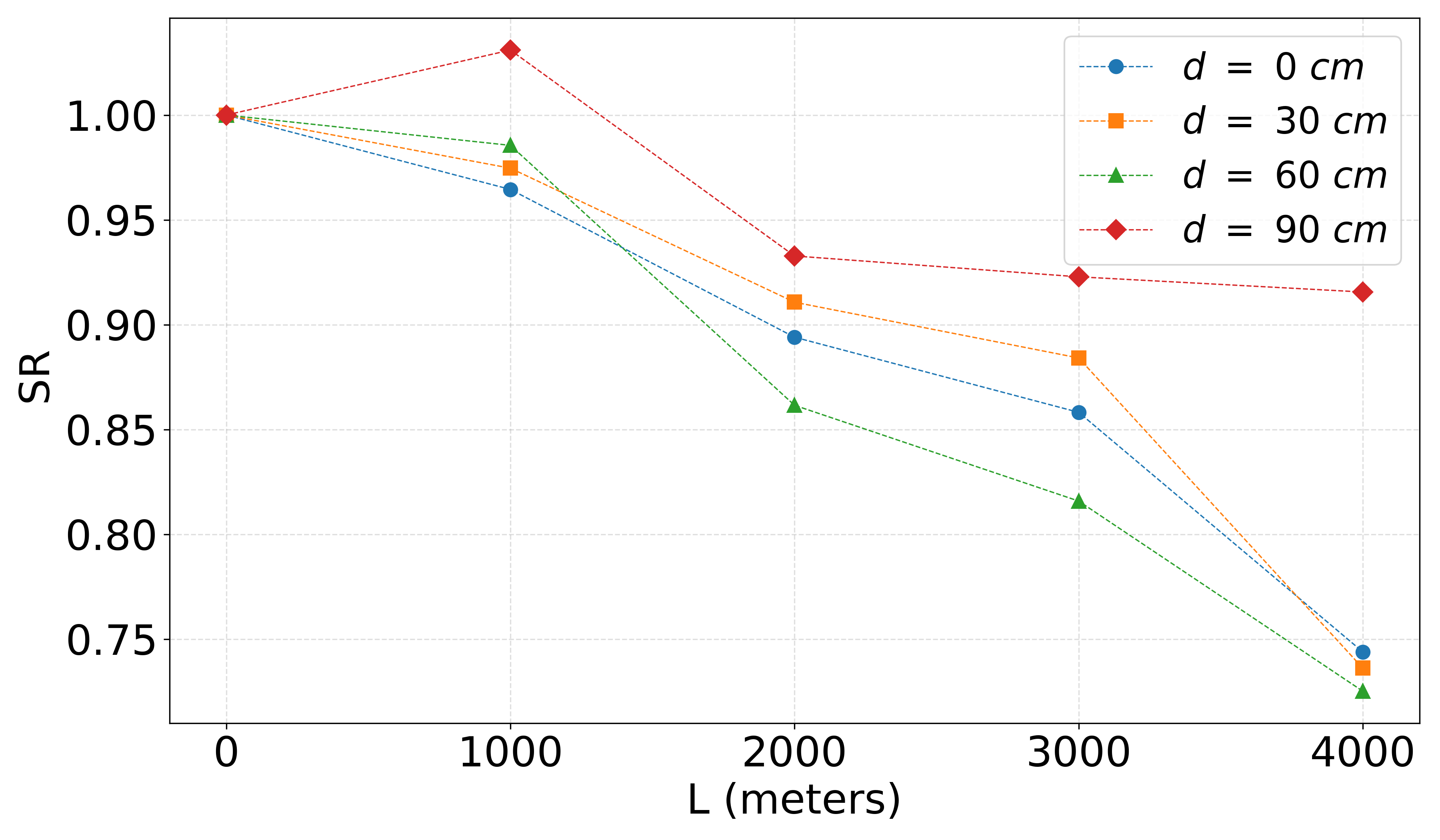}
\caption{\justifying Strehl Ratio (SR) calculated from intensity distributions as a function of the simulated turbulence propagation distance (from $0\text{ m}$ to $4000\text{ m}$): 
\textbf{(top)} Laguerre-Gaussian (LG) modes ($\ell \in \{1, \dots, 5\}$), displaying severe mathematical anomalies and non-monotonic peaks ($\text{SR} > 1$) driven by severe scintillation and localized hotspot formation; 
\textbf{(middle)} Hermite-Gaussian (HG) modes ($m=0, n \in \{1, \dots, 5\}$), showing a rapid decay as turbulence randomly redistributes energy away from the initial spatial structures; and 
\textbf{(bottom)} Airy beams, illustrating an enhanced and highly predictable stability sustained by their intrinsic self-healing properties under phase perturbations.}
\label{fig:SR}
\end{figure}

\subsubsection{Laguerre-Gaussian Modes}

Figure \ref{fig:SR} (top) presents the SR results for LG modes. Under strong turbulence, the characteristic azimuthal phase and the associated ring-shaped intensity profile undergo significant degradation. Energy is forced into the central "dark" region of the mode, eventually leading to the formation of complex speckle patterns. Notably, for certain realizations, LG modes can exhibit $SR > 1$. This is not indicative of beam focusing, but rather a manifestation of scintillation where a sharp, random speckle peak exceeds the relatively lower peak intensity of the original (unperturbed) ring distribution. The variance observed across different topological charges ($\ell$) and propagation distances ($L$) is attributed to the stochastic nature of the phase screens used to simulate the turbulent medium.

\subsubsection{Hermite-Gaussian Modes}

The results for HG modes, shown in Figure \ref{fig:SR} (middle), indicate that with the exception of the $m=1$ case in specific realizations, the SR values remain predominantly below unity. In a manner similar to LG modes, the energy is redistributed into the nodal lines (dark areas) of the HG structure. However, the energy scattering in HG modes tends to broaden the existing lobes rather than generating isolated speckles that surpass the reference peak intensity, as reflected in the lower SR values compared to LG modes.

\subsubsection{Airy Beams}

Figure \ref{fig:SR} (bottom) illustrates the SR performance for Airy beams. A clear trend of decreasing SR is observed as the propagation distance ($L$) in the turbulent medium increases. Nevertheless, Airy beams demonstrate a unique structural resilience. Due to the self-healing effect, the side lobes of the Airy distribution act as an energy reservoir, redistributing power toward the main lobe to reconstruct the peak intensity even after significant perturbation. Consistent with the NCC analysis, the SR data confirms that Airy beams are significantly more resilient to atmospheric turbulence than their Gaussian-based counterparts.

\subsection{Beam Width Broadening}

The Beam Width Broadening (BRO) factor evaluates the transverse expansion of the beams by calculating the beam width through the second-moment method (Equation \ref{eq:Second_moment}) for cases both with and without turbulence. The BRO method, as defined in Equation \ref{eq:BRO}, compares the reference beam (in free space) to its state after propagation through a turbulent medium. This metric quantifies the energy dispersion relative to the beam's centroid. In a vacuum or weak turbulent medium, we expect $BRO \approx 1$, indicating that the beam's energy distribution remains largely unperturbed. Under strong turbulence, the theory typically predicts $BRO > 1$, as cumulative phase distortions tend to spread the energy over a larger area. Finally, a third regime where $BRO < 1$ may occur, often associated with stochastic autofocalization or, more critically, modal fragmentation.

This method is perhaps the most subtle of the four metrics employed. The energy scattering induced by severe turbulence often creates localized hotspots, or speckles, that concentrate energy in small regions of the beam's profile. When a mode is considerably destroyed, the second-moment algorithm may "lock" onto a single intense speckle near the centroid, making the beam appear to have focused (autofocalization) when, in reality, its global structure has been decimated.

\begin{figure}[htbp]
    \centering
    \includegraphics[width=0.98\columnwidth]{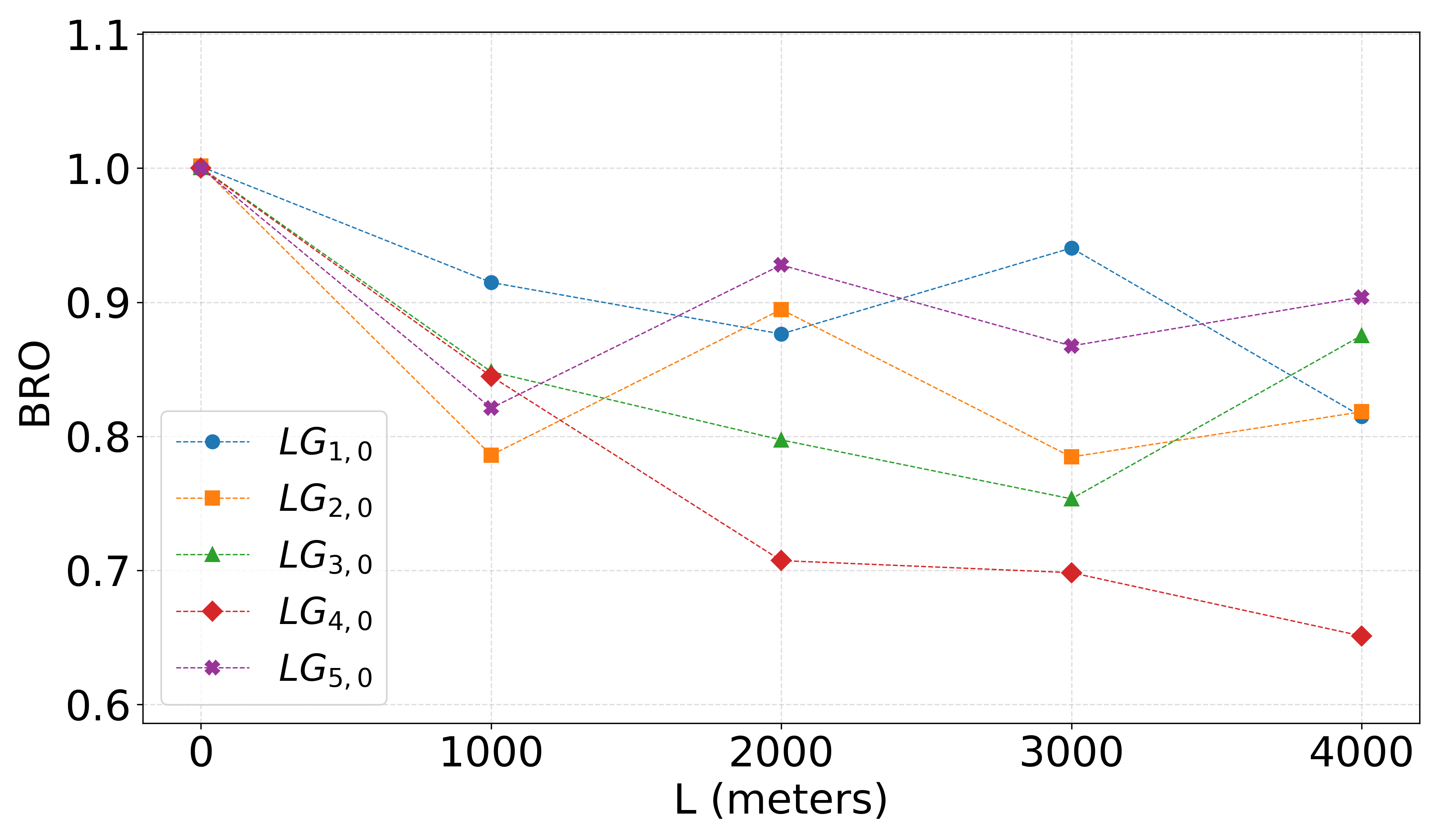}
    \vspace{0.2em}
    \includegraphics[width=0.98\columnwidth]{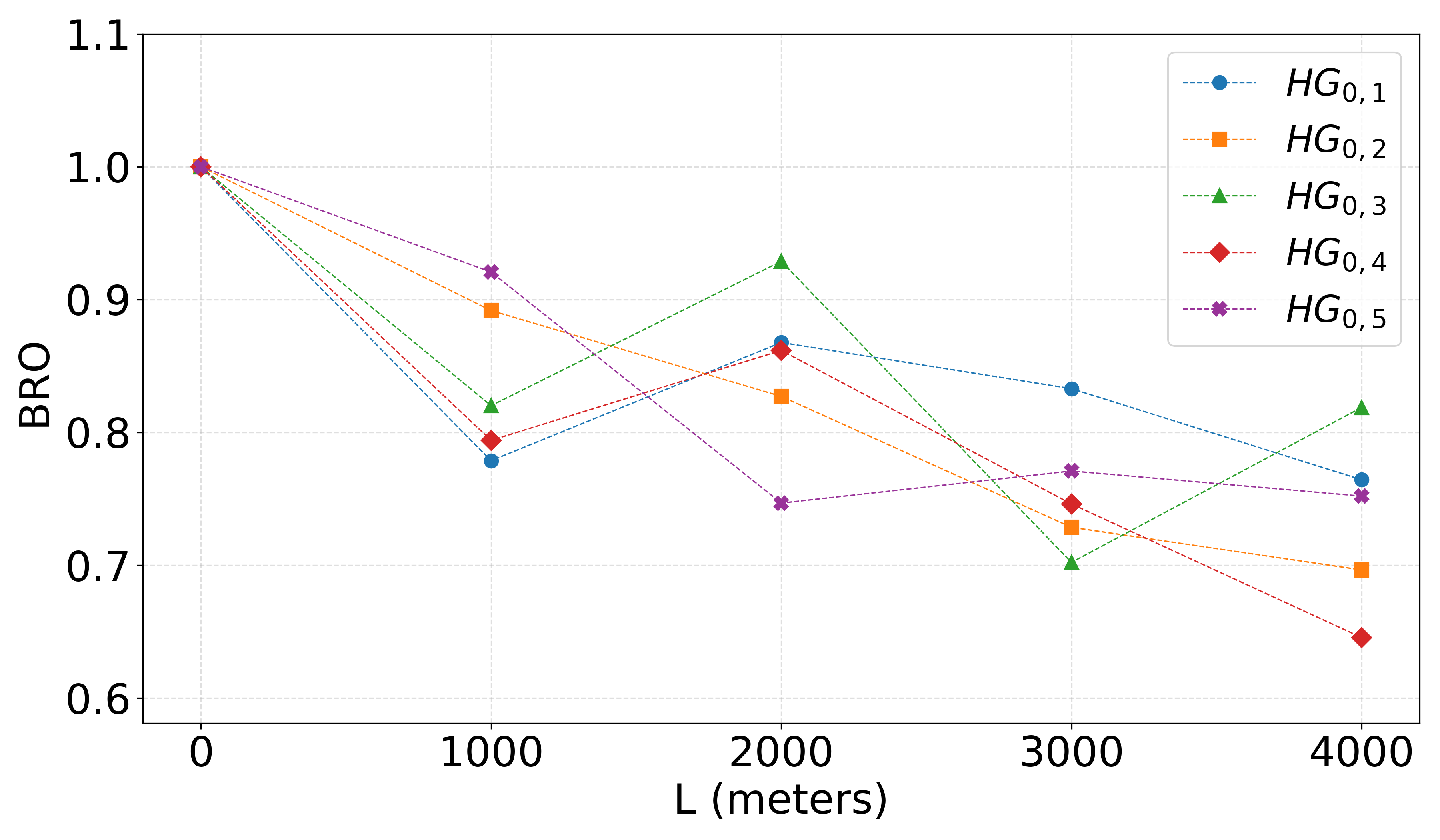}
    \vspace{0.2em}
    \includegraphics[width=0.98\columnwidth]{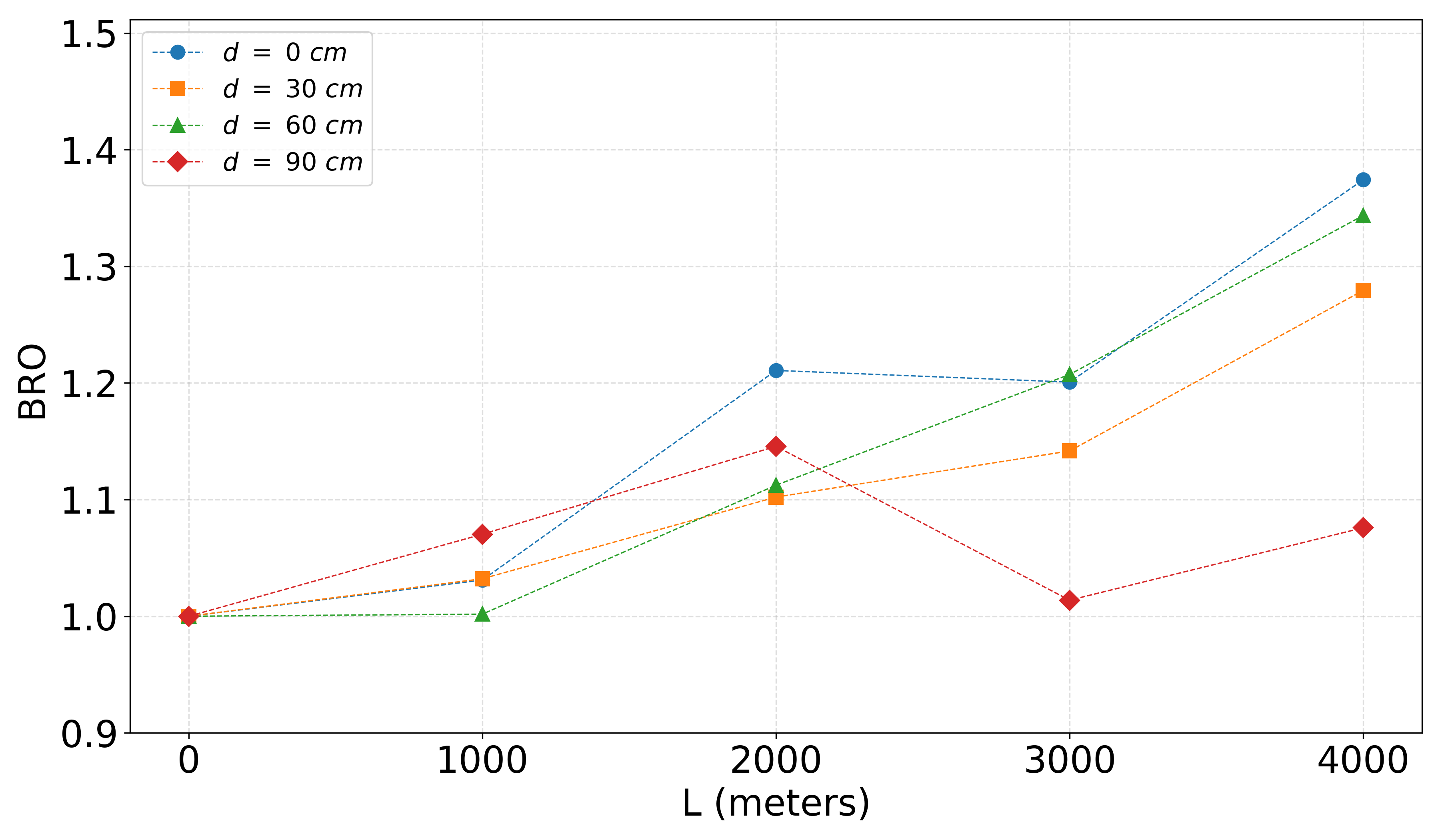}
\caption{\justifying Beam Width Broadening (BRO) factor calculated via the second-moment method as a function of the simulated turbulence propagation distance (from $0\text{ m}$ to $4000\text{ m}$): 
\textbf{(top)} Laguerre-Gaussian (LG) modes ($\ell \in \{1, \dots, 5\}$) and 
\textbf{(middle)} Hermite-Gaussian (HG) modes ($m=0, n \in \{1, \dots, 5\}$), both exhibiting an algorithmic paradox where severe modal fragmentation and speckle lock-on lead to non-physical beam-shrinking artifacts ($\text{BRO} < 1$); and 
\textbf{(bottom)} Airy beams, displaying a well-behaved, physical spatial broadening ($\text{BRO} > 1$) that reflects genuine energy diffusion, followed by structural stabilization driven by their self-healing dynamics.}
\label{fig:BRO}
\end{figure}

\subsubsection{Laguerre-Gaussian Modes}

Figure \ref{fig:BRO} (top) illustrates the results for LG beams. The data clearly shows that beam fragmentation into speckles concentrates the energy toward a single point, effectively destroying the original ring-shaped structure. This phenomenon explains why the second-moment method is not inherently suitable for evaluating LG modes under strong turbulence. While one would theoretically expect higher OAM modes to expand further—increasing the beam waist—the BRO metric captures the opposite effect: the beam appears to "become smaller." This paradox is a signature of modal collapse, where the metric fails to account for the total energy scattered into the background and focuses only on the fragmented centroid.

\subsubsection{Hermite-Gaussian Modes}

Figure \ref{fig:BRO} (middle) shows the results for HG beams. An identical problem to that observed in LG beams occurs here: fragmentation into speckles forces energy into the central dark areas or nodal lines, causing the mode to lose its definition. We observe a similar trend in the BRO factor, where values drop as the beam collapses into a "single fragment of light." This confirms that the BRO method is likewise unsuited for characterizing HG modes in high-turbulence regimes, as it misinterprets fragmentation as a reduction in beam width.

\subsubsection{Airy Beams}

Figure \ref{fig:BRO} (bottom) presents the BRO results for Airy beams, which align more closely with theoretical expectations. Here, the BRO grows consistently from 1 to approximately 1.4 as the propagation distance in the turbulent medium ($L$) increases. This indicates that the Airy beam maintains its structural integrity while undergoing diffusion. Furthermore, the self-healing behavior is evident as the BRO tends back toward unity as the camera distance increases. This suggests that during free propagation (the distance $d$), the beam is capable of recovering its main lobe intensity from its lateral energy reservoir, effectively mitigating the broadening induced by the preceding turbulence.

\subsection{Scintillation Index (SC)}

The Scintillation Index (SC) quantifies the intensity fluctuations induced by the propagation medium, as defined by the normalized variance of intensity in Equation \ref{eq:scintilation_index}. In this work, we employ the scintillation ratio ($SC_{ratio}$), defined in Equation \ref{eq:SC}, which compares the turbulent beam's fluctuations against those of the reference beam. This parameter illustrates how the intensity distribution is perturbed and scattered by the turbulent process.

In the limit of strong turbulence, the scintillation of the turbulent beam tends to unity ($\sigma^2_{T} \rightarrow 1$), leading to the saturation of the $SC_{ratio}$ as follows:
\begin{equation} \label{eq:SC_sat}
    SC_{sat} \approx \frac{1}{\sigma^2_0}.
\end{equation}
This theoretical limit defines the maximum relative scintillation reachable before saturation, which is intrinsically dependent on the stability of the reference beam ($\sigma^2_0$). In our analysis, the average $SC_{sat}$ is calculated for each reference measurement set to determine whether the system has reached the saturation regime. High SC values indicate a less reliable channel, as severe intensity fluctuations can lead to signal instability and fading in optical communication.

\begin{figure}[htbp]
    \centering
    \includegraphics[width=0.98\columnwidth]{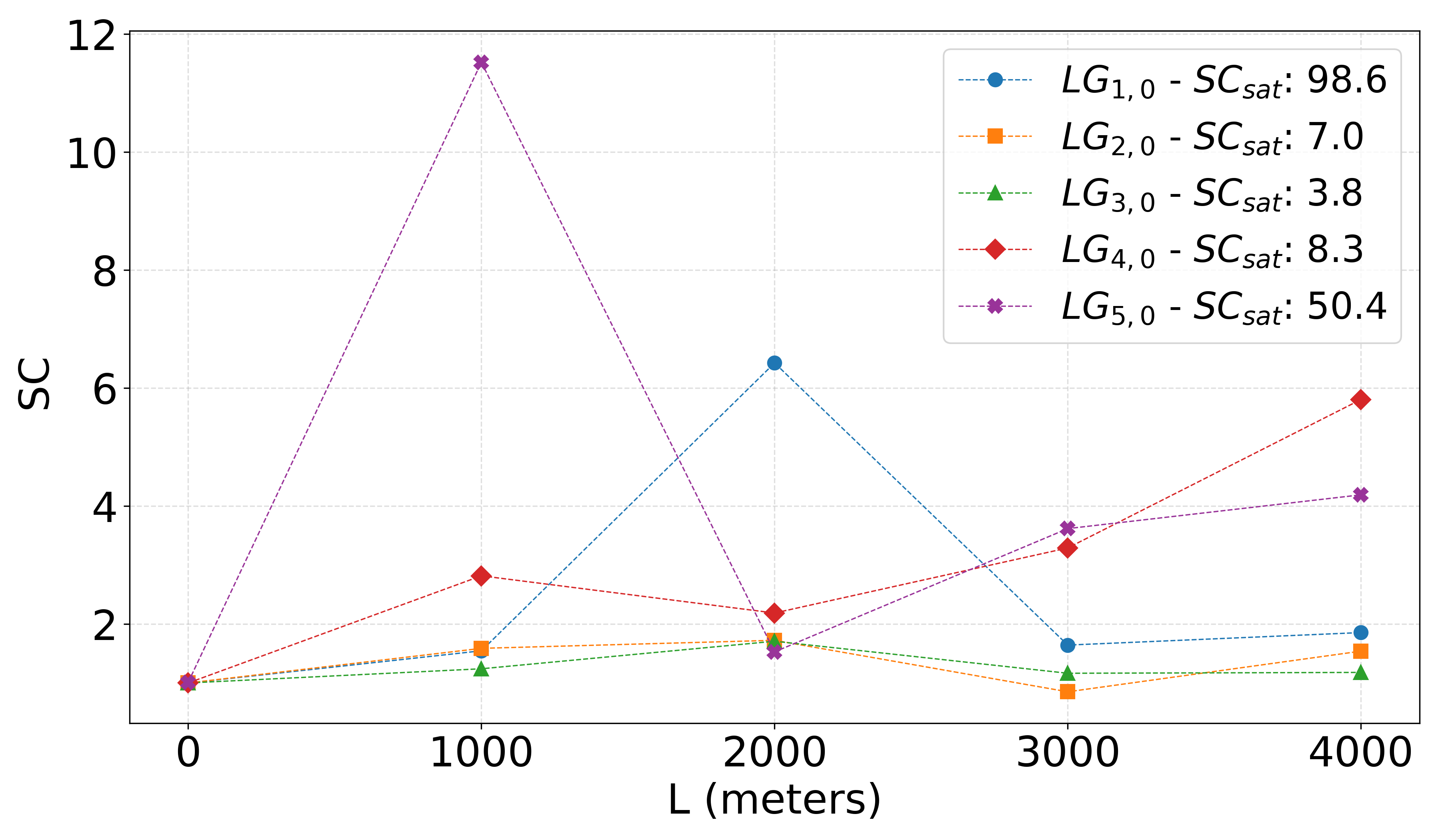}
    \vspace{0.2em}
    \includegraphics[width=0.98\columnwidth]{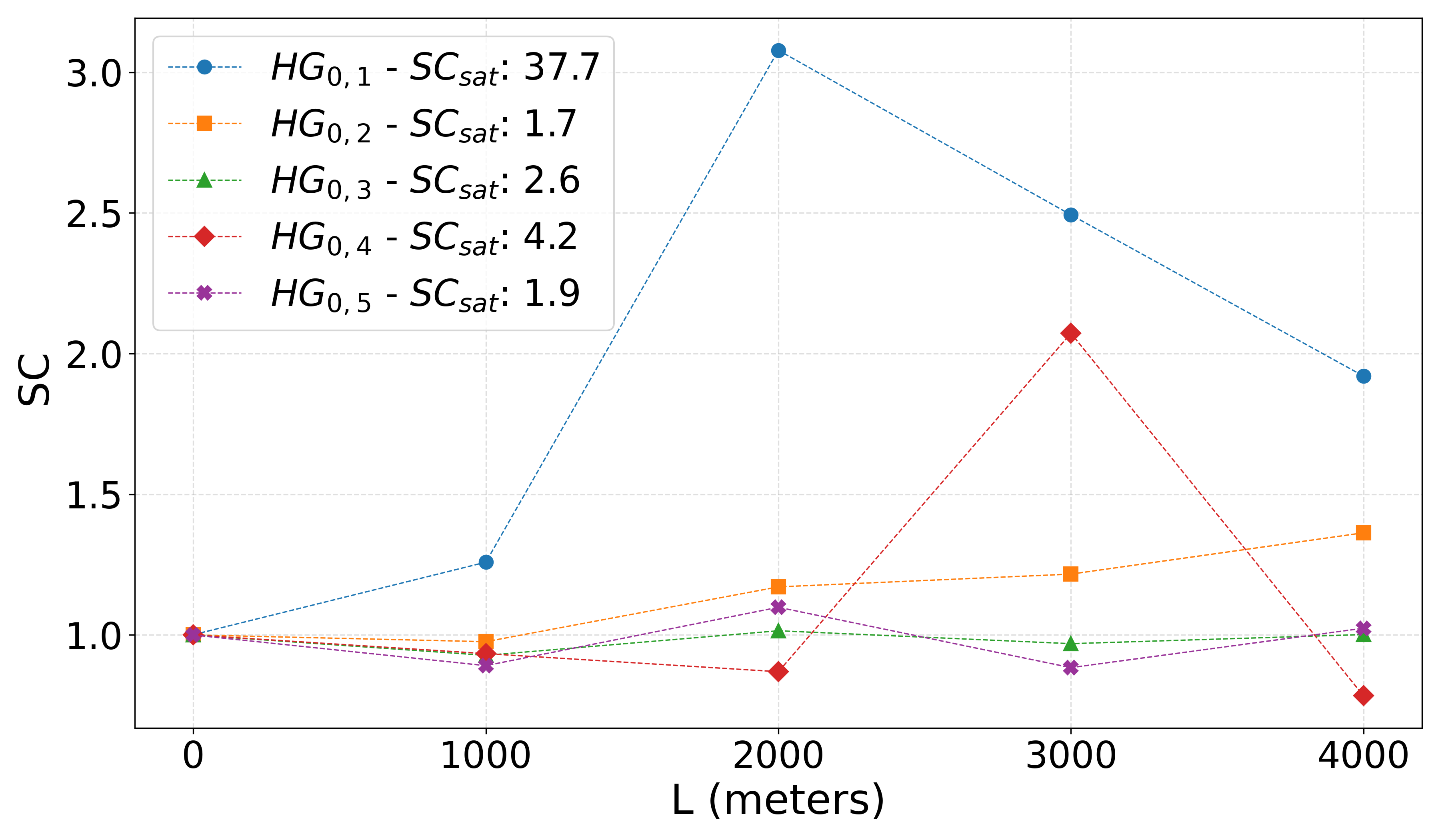}
    \vspace{0.2em}
    \includegraphics[width=0.98\columnwidth]{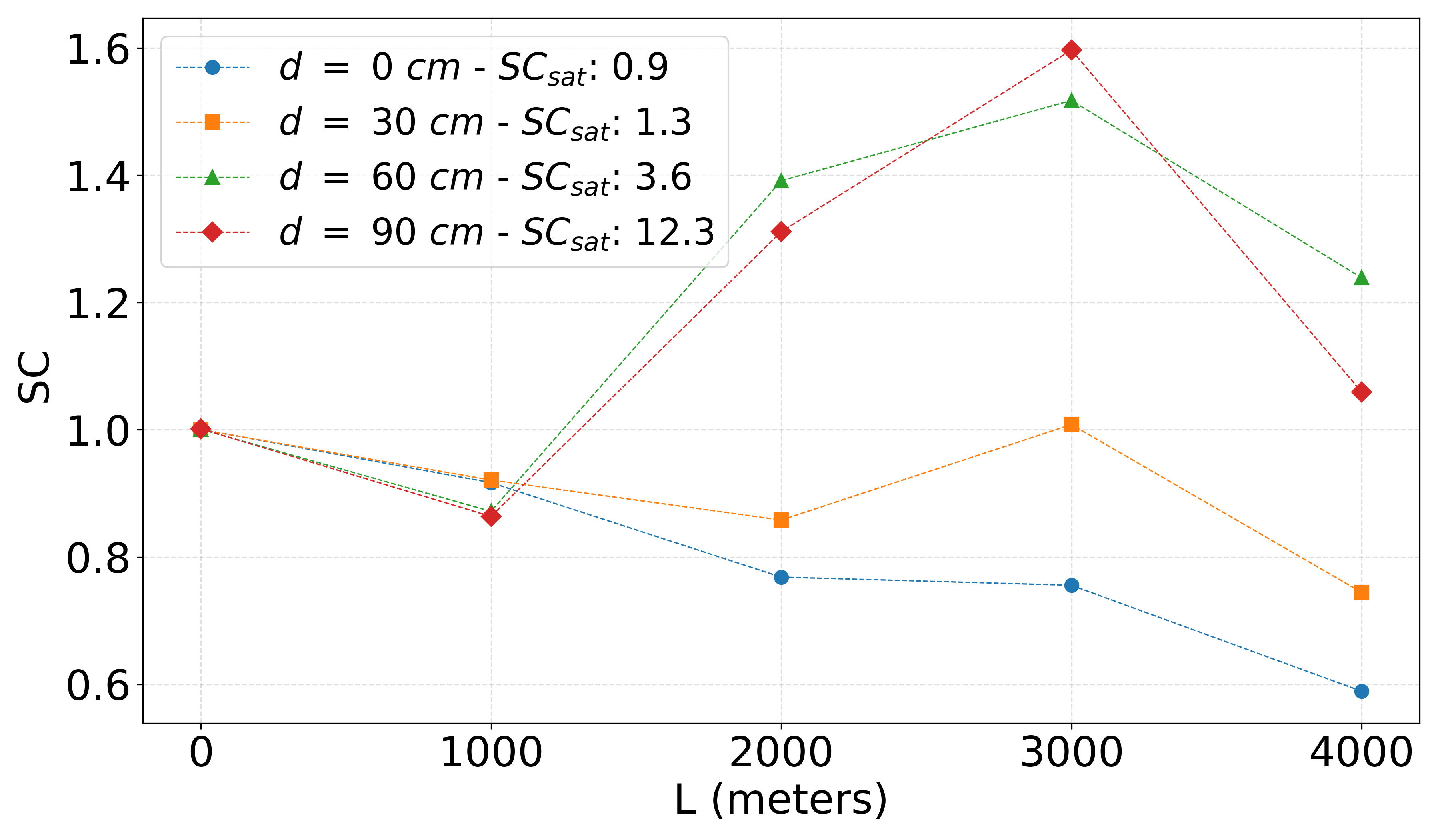}
\caption{\justifying Scintillation Index (SC) ratio quantifying intensity fluctuations under simulated atmospheric turbulence as a function of propagation distance (from $0\text{ m}$ to $4000\text{ m}$), with the respective saturation limits for each measurement set explicitly plotted for reference: 
\textbf{(top)} Laguerre-Gaussian (LG) modes ($\ell \in \{1, \dots, 5\}$), showing severe fluctuation scaling that approaches the saturation regime; 
\textbf{(middle)} Hermite-Gaussian (HG) modes ($m=0, n \in \{1, \dots, 5\}$), illustrating how higher-order spatial fracturing introduces spatial blurring that suppresses fast intensity spikes ($\text{SC} < 1$); and 
\textbf{(bottom)} Airy beams, demonstrating statistical mitigation of scintillation due to the continuous phase smoothing enabled by their distributed transverse profile.}
\label{fig:SC}
\end{figure}

\subsubsection{Laguerre-Gaussian Modes}

Figure \ref{fig:SC} (top) presents the results for LG beams. As expected, a general increase in the SC ratio is observed for nearly all measurements as the propagation distance in the turbulent medium ($L$) increases. Some realizations exhibit significant peaks, which are attributed to the stochastic nature of the turbulence holograms. Higher-order modes (larger $\ell$) are more severely affected than lower-order ones; since the beam waist—and consequently the beam area—increases with $\ell$, the interaction cross-section with the turbulent cells is larger, leading to more pronounced intensity spikes within the central "dark" region of the mode.

\subsubsection{Hermite-Gaussian Modes}

Figure \ref{fig:SC} (middle) shows the results for HG beams ($m=0, n \neq 0$). For the $n=1$ mode, the SC values frequently "explode" or exhibit high peaks, a behavior primarily driven by \textit{beam wander}, where small-scale turbulence deflects the entire beam centroid. However, for higher-order modes (e.g., $n=5$), a different phenomenon occurs: the light scatters into the nodal lines (dark areas), creating a significant spatial blurring effect. In this regime, the beam loses its modal definition and becomes a diffuse distribution of light. This homogenization of the intensity profile reduces the normalized variance relative to the high-contrast reference beam, resulting in $SC_{ratio} < 1$.

\subsubsection{Airy Beams}

Figure \ref{fig:SC} (bottom) illustrates the SC results for Airy beams. At shorter free-space propagation distances ($d$), the beam appears diffuse, and the contrast between the main and secondary lobes is diminished, similar to the blurring observed in HG modes. This diffused state spreads the intensity and leads to a reduction in the SC ratio ($SC < 1$). Nevertheless, as the propagation distance increases, the self-healing mechanism facilitates the recovery of the beam's structure. The energy from the secondary lobes is redistributed toward the main lobe, reconstructing the characteristic Airy profile and stabilizing the intensity fluctuations compared to the Gaussian-based counterparts.

\section{Conclusion}\label{sec:conclusion}

Laguerre–Gaussian beams possess a well-defined azimuthal phase structure and, under ideal conditions, carry orbital angular momentum associated with a helical wavefront. Atmospheric turbulence disrupts this helical phase by introducing random phase variations that differ across the cross section of the beam. As a result, the phase singularity at the center of the beam becomes distorted and the azimuthal phase symmetry is broken. The orbital angular momentum content is no longer confined to a single mode but becomes distributed across a spectrum of modes.

This process, often referred to as mode coupling or mode crosstalk, leads to a degradation of the modal purity of the beam. Although the total angular momentum is conserved in a statistical sense, its distribution among spatial modes fluctuates from realization to realization. Intensity-wise, the characteristic ring structure of the Laguerre-Gaussian beams becomes irregular, and eventually indistinguishable as the turbulence strength increases. These features make Laguerre–Gaussian beams particularly sensitive to atmospheric turbulence, especially in applications that rely on mode orthogonality or orbital angular momentum encoding.

Hermite–Gaussian beams are characterized by a Cartesian symmetry and a pattern of nodal lines where the intensity vanishes. These nodal structures arise from precise phase relationships between different transverse mode components of the field. Atmospheric turbulence perturbs these phase relationships, causing the nodal lines to shift, bend, or disappear altogether. The rectangular symmetry that defines Hermite–Gaussian modes is therefore progressively lost as the beam propagates through turbulence.

Unlike Laguerre–Gaussian beams, Hermite–Gaussian modes do not rely on azimuthal phase singularities but are still highly sensitive to phase distortions that break their separability along orthogonal axes. Turbulence-induced anisotropy can further enhance this degradation, preferentially distorting the beam along certain transverse directions. As a result, energy initially concentrated in a single Hermite–Gaussian mode spreads into multiple modes, leading to blurred intensity patterns and reduced contrast of the nodal structure.

Airy beams differ fundamentally from Gaussian-based modes in that they possess an asymmetric intensity profile and a phase structure that produces self-accelerating and self-healing behavior. When propagating through atmospheric turbulence, Airy beams also experience phase distortions and intensity fluctuations; however, their response is qualitatively different from that of the Laguerre-Gaussian and Hermite–Gaussian beams. Because the Airy beam’s structure is distributed over a wide range of spatial frequencies, localized phase perturbations do not immediately destroy its main lobe.

One notable consequence is that the primary intensity maximum of an Airy beam often remains identifiable even under moderate turbulence, although its trajectory becomes noisy and its acceleration less well defined. The side lobes, which play a crucial role in self-healing, are more severely affected and tend to degrade rapidly. Although turbulence ultimately disrupts the Airy structure, the beam can partially reconstruct itself after encountering localized distortions, giving it a degree of robustness compared to other structured beams.

In all cases, atmospheric turbulence transforms structured optical beams from deterministic mode solutions into statistical objects. The specific way in which a beam degrades depends on how its spatial structure and phase coherence are encoded. Laguerre–Gaussian beams are primarily affected by the destruction of the azimuthal phase order, Hermite–Gaussian beams by the distortion of nodal symmetry, and Airy beams by perturbations of their accelerating interference structure. Understanding these differences is essential for interpreting experimental images and for designing correlation-based measures to quantify the strength of the turbulence.

%----------------------------------
\begin{acknowledgments}
LCC and AZK acknowledge support from the National Council for Scientific and Technological Development (CNPq) through grants 308065/2022-0 (LCC), 422305/2023-5 (AZK), and 303502/2022-3 (AZK); the National Institute of Science and Technology for Applied Quantum Computing through CNPq grant 408884/2024-0, Goiás State Research Foundation (FAPEG) through grant 202510267001843, and São Paulo State Research Foundation (FAPESP) through grants 2025/23726-4, 2021/06823-5, and 2022/15036-0.
\end{acknowledgments}

\newpage
\newpage
\appendix
\begin{widetext}

%%%%%%%%%%%%%%%%%%%%%%%%%%%%%%%%%%%%%%%%%%%%%%%%%%%%%%%%%%%%%%%%
%%%%%%%%%%%%%%%%%%%%%%%%%%%%%%%%%%%%%%%%%%%%%%%%%%%%%%%%%%%%%%%%
%%%%%%%%%%%%%%%%%%%%%%%%%%%%%%%%%%%%%%%%%%%%%%%%%%%%%%%%%%%%%%%%
\section{Modes}
\label{app:modes}

Here we present explicit expressions for the structured beans considered in the main text. The Laguerre-Gaussian modes in cylindrical coordinates $(r,\phi,z)$ are given by
\begin{equation}
    \psi_{\mathrm{LG}}(r,\phi,z) = E_0\left(\frac{\sqrt{2}}{w}r\right)^{\ell}L^{\ell}_{p}\left(\frac{2r^{2}}{w^{2}}\right)\frac{w_0}{w(z)}e^{-i\xi_{p\ell}(z)}e^{i\frac{k}{2q(z)}r^{2}}e^{i\ell\phi}.
    \label{app:eq:LG}
\end{equation}
In this equation, $E_0$ is a constant intensity, omega is the frequency of light, $L_{p}^{l}$ the associated Laguerre polynomial, $w(z)=w_0\sqrt{1+z^2/z_0^2}$ is the beam size, $w_0$ is the beam size at the beam waist, $z_0 = \pi w_0^{2}/\lambda$ is the Rayleigh range, with $\lambda$ being the wavelength, $q=z-iz_0$ is the complex beam parameter, $\ell$ is the topological charge of the mode carrying Orbital Angular Momentum, and $\xi_{p\ell} = \left(2p+\abs{\ell}+1\right)\tan^{-1}(z/z_0)$ is the Gouy phase shift. 

The Hermite-Gaussian modes are expressed in Cartesian coordinates as
\begin{equation}
    \psi_{\mathrm{HG}}(x,y,z) = E_0H_n\left(\frac{\sqrt{2}}{w(z)}x\right)H_m\left(\frac{\sqrt{2}}{w(z)}y\right)\frac{w_0}{w(z)}e^{-i\xi_{mn}(z)}e^{i\frac{k}{2q(z)}r^{2}},
    \label{app:eq:HG}
\end{equation}
with $H_n$ denoting the Hermite polynomial and the Gouy phase shift is given by $\xi_{mn}(z) = \left(m+n+1\right)\tan^{-1}(z/z_0)$.

Finally, the Airy beam is written as $\psi_{\mathrm{Ai}} = \psi_x(x,z)\psi_y(y,z)$, with
\begin{equation}
    \psi_x(x,z) = \mathrm{Ai}\left(\frac{x}{x_0} - \frac{z^{2}}{4k^{2}x_0^{4}} + ia_x\frac{z}{kx_0^2}\right)\exp\left[a_x\left(\frac{x}{x_0} - \frac{z^{2}}{2k^{2}x_0^{4}}\right) + i\left(\frac{xz}{2kx_0^3} - \frac{z^{3}}{12k^{3}x_0^{6}} + \frac{a_x^2 z}{2kx_0^2}\right)\right],
    \label{app:eq:Ai}
\end{equation}
with $x_0$ being the characteristic transverse size of the Airy beam that sets the spacing of the Airy lobes, the width of the main lobe, and the curvature of the accelerating trajectory. $a_x$ is a dimensionless parameter introduced to make the Airy beam physically realizable (avoiding infinite energy). The function $\psi_y(y,z)$ is obtained from this by making $x\rightarrow y$, $x_0\rightarrow y_0$, and $a_x\rightarrow a_y$.

%%%%%%%%%%%%%%%%%%%%%%%%%%%%%%%%%%%%%%%%%%%%%%%%%%%%%%%%%%%%%%%%
%%%%%%%%%%%%%%%%%%%%%%%%%%%%%%%%%%%%%%%%%%%%%%%%%%%%%%%%%%%%%%%%
%%%%%%%%%%%%%%%%%%%%%%%%%%%%%%%%%%%%%%%%%%%%%%%%%%%%%%%%%%%%%%%%
\section{Discrete Intensity Matrices}
\label{app:discrete}

The CCD camera consists of $N_x \times N_y$ pixels. The measured images are represented as discrete matrices $I_0[i,j]$ and $I_T[i,j]$ for the reference and turbulent images, respectively. $i = 1,\dots,N_x$ and $j = 1,\dots,N_y$ label the pixel coordinates.

Each matrix element is proportional to the time-integrated optical intensity:
\begin{equation}
    I[i,j] \propto \int_0^{\tau} \abs{E(x_i,y_j,t)}^2 \, \dd t,
\end{equation}
where $E(x,y,t)$ is the optical field and $\tau$ is the camera exposure time.

Throughout this discussion, we assume that the exposure time is sufficiently long that the recorded turbulent image represents a statistical average over many realisations of the turbulent medium.

Before any comparison is performed, both images are normalised to remove trivial intensity fluctuations. Therefore, we write
\begin{equation}
    I[i,j] \;\longrightarrow\; \frac{I[i,j]}{\sum_{i,j} I[i,j]}.
\end{equation}

Now, centroid alignment is applied to remove global beam wander:
\begin{equation}
    x_c = \frac{\sum_{i,j} x_i I[i,j]}{\sum_{i,j} I[i,j]}, \quad
    y_c = \frac{\sum_{i,j} y_j I[i,j]}{\sum_{i,j} I[i,j]}.
\end{equation}
The images are shifted so that their centroids coincide. This ensures that the comparison isolates wavefront distortions and spatial structure degradation, rather than trivial translations.

Let us define the mean intensities as
\begin{equation}
    \bar{I}_0 = \frac{1}{N} \sum_{i,j} I_0[i,j], \quad
    \bar{I}_T = \frac{1}{N} \sum_{i,j} I_T[i,j],
\end{equation}
where $N = N_x N_y$ is the total number of pixels. The intensity fluctuations thus take the form
\begin{equation}
    \delta I_0[i,j] = I_0[i,j] - \bar{I}_0, \quad
    \delta I_T[i,j] = I_T[i,j] - \bar{I}_T.
\end{equation}
The normalised cross-correlation coefficient is then
\begin{equation}
    \mathcal{C}
    =
    \frac{
        \sum_{i,j} \delta I_0[i,j] \, \delta I_T[i,j]
    }{
        \sqrt{
        \left( \sum_{i,j} \delta I_0[i,j]^2 \right)
        \left( \sum_{i,j} \delta I_T[i,j]^2 \right)
        }
    }
\end{equation}

\end{widetext}

%-----------------------------------


\begin{thebibliography}{99}

%1
\bibitem{rsa1978}
R.~L. Rivest, A.~Shamir, e L.~Adleman,
\newblock "A method for obtaining digital signatures and public-key cryptosystems,"
\newblock {\em Communications of the ACM}, vol. 21, no. 2, pp. 120--126, 1978.

%2
\bibitem{Stinson2023} D. R. Stinson and M. Paterson, \emph{Cryptography: Theory and Practice} (CRC Press, 2023).

%3
\bibitem{Menezes1996} A. J. Menezes, P. C. van Oorschot, and S. A. Vanstone, \emph{Handbook of Applied Cryptography} (CRC Press, 1996).

%4

\bibitem{BB84}
C.~H. Bennett and G.~Brassard.
\newblock Quantum cryptography: Public key distribution and coin tossing.
\newblock In {\em Proceedings of IEEE International Conference on Computers,
  Systems and Signal Processing}, volume 175, page ~8, 1984.

%5
\bibitem{E91}
A.~K. Ekert.
\newblock Quantum cryptography based on {Bell’s} theorem.
\newblock {\em Physical Review Letters}, 67(6):661--663, 1991.

%6

\bibitem{Gisin2002} N. Gisin, G. Ribordy, W. Tittel, and H. Zbinden, Quantum cryptography, Rev. Mod. Phys. \textbf{74}, 145 (2002).


%7
\bibitem{Scarani2009} V. Scarani, H. Bechmann-Pasquinucci, N. Cerf, M. Du{\v s}ek, N. L{\"u}tkenhaus, and M. Peev, The security of practical quantum key distribution, Rev. Mod. Phys. \textbf{81}, 1301 (2009).
%8
\bibitem{Agrawal2002}
G. P. Agrawal,
\emph{Fiber-Optic Communication Systems},
3rd ed.
(John Wiley \& Sons, New York, 2002).
%9


\bibitem{Flamini2019}
F. Flamini, N. Spagnolo, and F. Sciarrino,
Photonic quantum information processing: a review,
Rep. Prog. Phys. \textbf{82}, 016001 (2019).


%10
\bibitem{allen1992} L. Allen, M. W. Beijersbergen, R. J. C. Spreeuw, and J. P. Woerdman, Orbital angular momentum of light and the transformation of Laguerre-Gaussian laser modes, Phys. Rev. A \textbf{45}, 8185 (1992).

%11
\bibitem{mair2001} A. Mair, A. Vaziri, G. Weihs, and A. Zeilinger, Entanglement of the orbital angular momentum states of photons, Nature \textbf{412}, 313 (2001).

%12
\bibitem{boyd2005} M. N. O’Sullivan-Hale, I. A. Khan, R. W. Boyd, and J. C. Howell, Pixel entanglement: Fine-grained spatial correlations for secure quantum communication, Phys. Rev. Lett. \textbf{94}, 220501 (2005).

%13
\bibitem{Kaushal2017}
H.~Kaushal and G.~Kaddoum,
\newblock Optical communication in space: Challenges and mitigation techniques,
\newblock {\em IEEE Commun. Surveys Tuts.} \textbf{19}, 57 (2017).
%14
\bibitem{saleh_livro}
B.~E.~A. Saleh e M.~C. Teich,
\emph{Fundamentals of Photonics},
2\textsuperscript{a}~ed.
Hoboken, NJ: John Wiley \& Sons, 2007.

%15
\bibitem{Erhard2020}
M.~Erhard, M.~Krenn, and A.~Zeilinger,
\newblock Advances in high-dimensional quantum information,
\newblock {\em Nat. Rev. Phys.} \textbf{2}, 365--381 (2020).

% 16
\bibitem{rodenburg2012} B. Rodenburg, M. P. J. Lavery, M. Malik, M. N. O'Sullivan, M. Mirhosseini, D. J. Robertson, M. Padgett, and R. W. Boyd, Influence of atmospheric turbulence on states of orbital angular momentum of light, New J. Phys. \textbf{14}, 033014 (2012).

%17
\bibitem{paterson2005} C. Paterson, Atmospheric turbulence and orbital angular momentum of light, Phys. Rev. Lett. \textbf{94}, 153901 (2005).

%18
\bibitem{OAM_turbo} A. H. Ibrahim, F. S. Roux, M. McLaren, T. Konrad, and A. Forbes, Orbital-angular-momentum entanglement in turbulence, Phys. Rev. A \textbf{88}, 012312 (2013).

%19

\bibitem{siviloglou2007} G. A. Siviloglou, J. Broky, A. Dogariu, and D. N. Christodoulides, Observation of accelerating Airy beams, Phys. Rev. Lett. \textbf{99}, 213901 (2007).

%20
\bibitem{gu2010} Y. Gu and G. Gbur, Scintillation of pseudo-partially coherent Airy beams in atmospheric turbulence, Opt. Lett. \textbf{35}, 3456 (2010).

%21

\bibitem{klug2023robust}
A.~Klug, C.~Peters, and A.~Forbes, 
\emph{Robust structured light in atmospheric turbulence}, 
Adv. Photonics \textbf{5}, 016006 (2023).

%22
\bibitem{Goodman2005}
J. W. Goodman,
\emph{Introduction to Fourier Optics},
3rd ed.
(Roberts \& Company Publishers, Greenwood Village, CO, 2005).

%23
\bibitem{fried1966statistics} D. L. Fried, Statistics of a geometric representation of wavefront distortion, J. Opt. Soc. Am. \textbf{56}, 1372 (1966).
%24
\bibitem{SLM_artigoOriginal_tecnica} V. Arriz\'on, U. Ruiz, R. Rosas, and D. S\'anchez-de-la-Llave, Pixelated phase computer holograms for the accurate encoding of scalar complex fields, J. Opt. Soc. Am. A \textbf{24}, 3500 (2007).
%25

\bibitem{Lewis1995}
J. P. Lewis,
Fast normalized cross-correlation,
Vision Interface \textbf{10}, 120--123 (1995).

%26


\bibitem{Andrews2005} L. C. Andrews and Ronald L. Phillips, \emph{Laser Beam Propagation through Random Media} (SPIE--The International Society for Optical Engineering, 2005).
%27
\bibitem{Tatarski2016} V I Tatarski and R. Silverman, \emph{Wave Propagation in a Turbulent Medium} ( Dover Publications, 2016).
%28
\bibitem{Tatarski1971} V. Tatarski, \emph{The Effect of the Turbulent Atmosphere on Wave Propagation} (Israel Program for Scientific Translations, Jerusalem, 1971).
%29
\bibitem{Ishimaru1978} A. Ishimaru, \emph{Wave Propagation and Scattering in Random Media} (Academic Press, NY, 1978).
%30
\bibitem{Andrews2001} L. Andrews, R. Phillips, and C. Hopen, \emph{Laser Beam Scintillation with Applications} (SPIE Press, Washington, 2001).
%31
\bibitem{Mandel1995} L. Mandel and E. Wolf, \emph{Optical Coherence and Quantum Optics} (Cambridge University Press, Cambridge, 1995).
%32

\bibitem{Schulze2013}
C. Schulze, A. Dudley, D. Flamm,
M. Duparré, and A. Forbes,
Probing the orbital angular momentum spectrum of optical beams through atmospheric turbulence,
New J. Phys. \textbf{15}, 073025 (2013).























\end{thebibliography}
\end{document}